\documentclass[12pt,preprint]{aastex}

\newcommand{\var}{\varpi}
\newcommand{\tha}{\theta}
\newcommand{\alp}{\alpha}
\newcommand{\beq}{\begin{equation}}
\newcommand{\eeq}{\end{equation}}
\newcommand{\bea}{\begin{array}}
\newcommand{\eea}{\end{array}}
\newcommand{\dw}{\Delta \varpi}

\slugcomment{Submitted to ApJ}

\shorttitle{Apsidal Resonance in Planetary Systems }
\shortauthors{ZHOU and SUN}

\begin{document}

\title{Occurrence and Stability of Apsidal Resonance \\
in Multiple Planetary Systems}

\author{Ji-Lin Zhou,  Yi-Sui Sun }
\affil{Department of Astronomy , Nanjing University,
    Nanjing 210093, China}
\email{(zhoujl@nju.edu.cn and sunys@nju.edu.cn)}


\begin{abstract}

With the help of the Laplace-Lagrange solution of the secular perturbation theory in a double-planet system,
 we study the occurrence and the stability
of apsidal secular resonance between the two planets. The explicit
criteria to predict whether two planets are in apsidal resonance
is derived, which shows the occurrence of the apsidal resonance
depends only on the mass ratio ($m_1/m_2$), semi-major axis
ratio($a_1/a_2$), initial eccentricity ratio ($e_{10}/e_{20}$),
and the initial relative apsidal longitude
($\varpi_{20}-\varpi_{10}$) between the two planets. The
probability of two planets falling in apsidal resonance is given
in the initial element space. We verify the criteria with
numerical integrations for the  HD12661 system, and find
 they give good predictions except at the boundary of the criteria or
 when the planet eccentricities are too large.
  The nonlinear stability of the two
planets in HD12661 system are studied by calculating the Lyapunov
exponents of their orbits in a general three-body model.
 We find that,  two planets in large eccentricity orbits
  could be stable only when they are in aligned apsidal resonance.
 When the planets are migrated under the planet-disk interactions,
 for more than half of the studied cases, the configurations of the apsidal resonances
 are preserved. We find the two planets of HD12661 system could be in aligned
resonance thus more stable provided they have
$\Omega_2-\Omega_1\approx 180^o$. The applications of the criteria
to the other multiple planetary systems are discussed.

\end{abstract}



\section{INTRODUCTION}

The detecting of extrasolar planetary systems reveals fruitful
results during the past years. More than 100 extrasolar planets
has been inferred by the Doppler radial velocity measurements to
the solar-type stars \citep{cc03}, among them 10 multiple-planet
systems are confirmed.
  For a multiple-planet system, the dynamical stability of the system under planetary interaction is an important issue
 concerning the dynamical evolution as well as the possible existence of habitable zone of the system.

There are many effects which can affect the stability of a multiple-planet system. For the orbits of planets with
small or modest eccentricities and inclinations, mean motion resonances between planets
can sometimes lead to stable configurations. Another effect is the secular resonance
between the planets. An apsidal resonance occurs
when the relative apsidal longitudes  of the two orbits $\dw$ librates about $0$ (aligned resonance)
and $\pi$ (anti-aligned resonance) during the evolution.
Due to the aligned apsidal resonance,
the two planets on elliptic orbits can greatly reduce the possibility of
 close encounters, thus
it is believed that aligned apsidal  resonance can stabilize the interacting planets.

For the 10 multiple planetary systems observed to date (GJ876,
47UMa, HD82943, HD12661, HD168443, HD37124, HD38529, HD74156,
Upsilon Andromedae, and 55 Cancri), the best fit orbital
parameters inferred from the radial velocity observations show 6
pairs of planets could be in apsidal resonance:
HD82943\citep{gm01}, Upsilon
Andromedae c and d(Chiang, Tabachnik, \& Tremaine 2001),
 GJ876\citep{lp02}, 47UMa
(Laughlin, Chambers, \& Fischer 2002),
HD12661\citep{gm03,lp03}, and 55 Cancri b and c\citep{ji03}.
Ubiquitous as it is, the apsidal resonance  phenomenon is worthy
to be studied in detail.
 In this paper, we are interested when the apsidal resonance occurs and
whether it really leads to a stable configuration between planets, since an anti-aligned resonance
could lead to close encounters between orbits with large eccentricities .

For the occurrence of apsidal
resonance, Laughlin et al. (2002) 
 gives a criterion  based on the Laplace-Lagrange solution of the secular perturbation system.
 In this paper, the criterion is represented to a more explicit form in section II.
 The probability that the apsidal resonance happens is also derived according to the criteria.
 In section III, we study the stability of orbits in apsidal resonances
 by calculating the largest Lyapunov exponents of the orbits with a general three-body model.
 The behavior of orbits in migration are studied with a torqued three-body model.
The conclusions and the applications of the criteria to
 the other multiple planetary systems are discussed in the final section.

\section{LOCATIONS OF APSIDAL RESONANCE}

In this section we derive the explicit criteria under which the apsidal
secular resonance may occur. To that aim, the linear secular perturbation theory is employed
for two interacting planets
 under the attraction of the host star. For a planetary system with two planets,
  hereafter we denote all the quantities of the host star, the inner and outer planets
 with subscripts ``0", ``1" and ``2", respectively.
  So the three bodies have masses $m_0,m_1,m_2$,  respectively, where
 $m_1, m_2 \ll m_0$. In the present study
 we address the coplanar problem only,
 so the two planets are on the orbits with osculating orbital elements
 $(a_1,e_1,\var_1,M_1)$ and $(a_2,e_2,\var_2,M_2)$, respectively, where $a,e,\var,M$ are
 the semi-major axis, eccentricity, longitude of pericenter and  mean anomaly of the orbit, respectively.
 We adopt the commonly used unit system, i.e., the mass unit is the solar mass,
 the length unit is 1AU, and the time unit is 1yr/($2\pi$).

\subsection{Linear secular perturbation theory revisited }

We start with the linear secular theory following  Murray and Dermott (1999). 
For the coplanar case, the disturbing function for the motions of
planets $m_1$ and $m_2$ are given as:

\begin{equation}
\begin{array}{cl}
  R_1= & n_1a_1^2~ [\frac12 A_{11} e_1^2+A_{12} e_1e_2 \cos (\var_1-\var_2)]  \\
  R_2= & n_2a_2^2~ [\frac12 A_{22} e_2^2+A_{21} e_1e_2 \cos (\var_1-\var_2)]
\end{array}
\label{a1}
\end{equation}
where $n_1,n_2$ are the mean motion of planets $m_1$ and $m_2$, respectively;
 $A_{ij}$ are elements of matrix given by
\begin{equation}
\left(
\begin{array}{cc}
  A_{11} & A_{12} \\
  A_{21} & A_{22}
\end{array}
\right) = \left(
\begin{array}{cc}
  c_1 & -c_0c_1 \\
  -c_0c_2 & c_2
\end{array}
\right)
\label{a2}
\end{equation}
where $c_k>0 ~(k=0,1,2)$ are functions of $a_1,a_2,m_0,m_1,m_2$ defined as
\beq
\begin{array}{rl}
  c_0=& b^{(2)}_{\frac32}(\alp)/b^{(1)}_{\frac32}(\alp)\approx  \frac54 \alp (1-\frac{1}{8} \alp^2 ) \\
  c_1=& \frac14 n_1\frac{m_2}{m_0+m_1}\alp^2 b^{(1)}_{\frac32}(\alp) \\
  c_2=& \frac14 n_2\frac{m_1}{m_0+m_2}\alp b^{(1)}_{\frac32}(\alp)
\end{array}
\label{a3}
\eeq
with $b^{(i)}_{\frac32}(\alp) (i=1,2)$ being the Laplace coefficients, and $\alp=a_1/a_2<1$.
Fig.1 shows the approximation of $c_0$ by the above formula. Quantitatively,  the error of approximation is
less than $1\%$ for $\alp< 0.66$, and less than $5\%$ for $\alp< 0.91$. So the approximation is quite well for the
study of the planetary system. Moreover, we define
\beq
\xi =\frac{c_2}{c_1}=\frac{1}{\alp}\frac{n_2m_1(m_0+m_1)}{n_1 m_2 (m_0+m_2)} \approx q \alp^{1/2},
\label{a3-8}
\eeq
with $q=m_1/m_2$ and the terms with orders of $O(\frac{m_1}{m_0}),O(\frac{m_2}{m_0})$ or higher
are neglected in the above approximation, since $\frac{m_1}{m_0},\frac{m_2}{m_0}\sim 10^{-3}$ in the planetary systems.
Denote $g_1,g_2$ as the two eigenvalues of matrix (\ref{a2}), and the corresponding eigenvectors are
$S_i\left( \bea{c} \cos \tha_{i}  \\ \sin \tha_{i} \eea \right)$ , where $\tha_i \in (-\pi/2, \pi/2)$ and
\beq
  \cos \tha_i=\frac{c_0c_1}{\sqrt{(c_1-g_i)^2+c_0^2c_1^2}},~
  \sin \tha_i=\frac{c_1-g_i}{\sqrt{(c_1-g_i)^2+c_0^2c_1^2}}, ~~(i=1,2)
\label{a3-2}
\eeq
with
\beq
\bea{l}
g_1=\frac{1}{2} [(c_1+c_2)+\sqrt{(c_1-c_2)^2+4c_0^2c_1c_2}] , \\
g_2=\frac{1}{2} [(c_1+c_2)-\sqrt{(c_1-c_2)^2+4c_0^2c_1c_2}] .
\eea
\label{a3-3}
\eeq
Define
\beq
\bea{ll}
\rho_1 \equiv \tan \tha_1 & = \frac{1}{2c_0}(1-\xi - \sqrt{(1-\xi)^2+4c_0^2 \xi}), \\
\rho_2 \equiv \tan \tha_2 & = \frac{1}{2c_0}( 1-\xi + \sqrt{(1-\xi)^2+4c_0^2 \xi}).
\eea
\label{a8}
\eeq
So $\rho_1 < 0$ and $\rho_2 > 0$, or $ -\frac{\pi}{2}<\tha_1<0<\tha_2<\frac{\pi}{2}$.
The scaling factor $S_i$ (i=1,2) can be expressed in terms of initial eccentricities $e_{10}$, $e_{20}$
and $\dw_0=\varpi_{20}-\varpi_{10}$:
\beq
\bea{l}
S_1= \frac{[ \rho_2^2e_{10}^2-2\rho_2 e_{10}e_{20} \cos \dw_0 +e_{20}^2]^{1/2}}{|\rho_1-\rho_2|\cos \tha_1}
\equiv \frac{F}{|\rho_1-\rho_2|\cos \tha_1}, \\
S_2= \frac{[ \rho_1^2e_{10}^2-2\rho_1 e_{10}e_{20} \cos \dw_0 +e_{20}^2]^{1/2}}{|\rho_1-\rho_2|\cos \tha_2}
\equiv \frac{G}{|\rho_1-\rho_2|\cos \tha_2}.
\eea
\label{s1s2}
\eeq

The secular system with disturbing functions (\ref{a1}) is integrable and the solutions can be written as:
\beq
\bea{cl}
e_1 & =\frac{1}{|\rho_1-\rho_2|}[F^2+2FG\cos \Delta \psi +G^2]^{1/2} \\
e_2 & =\frac{1}{|\rho_1-\rho_2|}[\rho_1^2 F^2+2 \rho_1\rho_2 FG\cos \Delta \psi +\rho_2^2 G^2]^{1/2}, \\
e_1e_2\sin \dw & = -\frac{1}{\rho_1-\rho_2} FG \sin \Delta \psi  \\
e_1e_2\cos \dw & = \frac{1}{(\rho_1-\rho_2)^2}[\rho_1 F^2+(\rho_1+\rho_2) FG \cos \Delta \psi +\rho_2G^2].
\label{e1e2}
\eea
\eeq
where $ \Delta \psi=\psi_2-\psi_1 =(g_2 t+\beta_2)-(g_1 t+\beta_1)$, with $t$ the time and
 $\beta_1,\beta_2$  given by:
\beq
\bea{ll}
\sin\beta_1=\frac{1}{F} ( h_{10} \rho_2 - h_{20}),  &
\sin\beta_2=-\frac{1}{G} ( h_{10} \rho_1 - h_{20}), \\
\cos\beta_1=\frac{1}{F} ( k_{10} \rho_2 - k_{20}) ,  &
\cos\beta_2=-\frac{1}{G} ( k_{10} \rho_1 - k_{20}).
\eea
\eeq
where $h_{i0}=e_{i0} \sin \varpi_{i0}$, $k_{i0}=e_{i0} \cos \varpi_{i0}$, $(i=1,2)$. From (\ref{e1e2}),
it's easy to verify that the evolution of $e_1,e_2$ obeys an integral:
\beq
 \frac{e_1^2}{A_{12}}+ \frac{e_2^2}{A_{21}}=D,
 \label{dd}
\eeq
where $D$ is a constant which depends only on the initial parameters.
Moreover, from (\ref{s1s2}) and (\ref{e1e2}), the maximum of $e_1$ and minimum of $e_2$ occur at $\cos \Delta \psi=1$
(as $\rho_1<0$), with values:
\beq
e_{1max}=\frac{F+G}{|\rho_1-\rho_2|}, ~~
e_{2min}=\frac{|\rho_1 F+\rho_2 G|}{|\rho_1-\rho_2|}.
\label{emax}
\eeq
The minimum of $e_1$ and maximum of $e_2$ occur at $\cos \Delta \psi=-1$, with
\beq
e_{1min}=\frac{|F-G|}{|\rho_1-\rho_2|}, ~~
e_{2max}=\frac{\rho_2 G-\rho_1 F}{|\rho_1-\rho_2|}.
\label{emin}
\eeq
Thus, we can obtain the maximum excursions of $e_1$ and $e_2$
 for any given $e_{10},e_{20}, \dw_0$ as follows:
\beq
\Delta e_1=\frac{(F+G)-|F-G|}{|\rho_1-\rho_2|}, ~~
\Delta e_2=\frac{(\rho_2 G-\rho_1 F)-|\rho_1 F+\rho_2 G|}{|\rho_1-\rho_2|}
\label{de}
\eeq

\subsection{The explicit criteria}


 With the help of the last equation of (\ref{e1e2}),  the criterion in Laughlin et al. (2002) for the apsidal resonance
 can be expressed as
\beq
 S= \left|
 \frac{(\rho_1+\rho_2) FG}{\rho_1 F^2+ \rho_2 G^2}
  \right|<1
  \label{a6}
\eeq
Since when $S<1$, the values of $\dw$ can not reach $\pi/2$ or $3\pi/2$ (thus  $\cos \dw \neq 0$), so it
must librate about $0$ or $\pi$. On the contrary, when
$S>1$, it is possible that $\dw$ will reach $\pi/2$ or $3\pi/2$ , thus it will circulate in $[0,2\pi]$.

The equation (\ref{a6}) equivalent to, after some algebra manipulations:
\beq
 \frac{F}{G} > \max(1,-\rho_1/\rho_2), ~~ {\rm or},~~
 0<\frac{F}{G} < \min(1,-\rho_1/\rho_2),
\label{a16}
\eeq
In view of (\ref{s1s2}), the above relations are equivalent to :
\beq
 \frac{e_{20}}{e_{10}} < \frac{2\rho_1 \rho_2}{\rho_1+ \rho_2} \cos \dw_0,
 \label{a17-3}
\eeq
or,
\beq
 \frac{e_{20}}{e_{10}} > \frac{\rho_1+ \rho_2}{2} \frac{1}{\cos \dw_0}>0.
\label{a17-4}
\eeq
By substituting  (\ref{a3}) (\ref{a3-8}) and (\ref{a8}) into the above expressions, we finally obtain:
\beq
 \frac{e_{20}}{e_{10}} < -\frac{5}{2}  \frac{q\alp^{3/2}(1-\frac18 \alp^2)}{1-q\alp^{1/2}} \cos \dw_0,
 \label{a17-1}
\eeq
or,
\beq
 \frac{e_{20}}{e_{10}} > \frac25 \frac{1-q\alp^{1/2}}{\alp(1-\frac18 \alp^2)} \frac{1}{\cos \dw_0}>0.
\label{a17-2}
\eeq
These  are the explicit criteria for the occurrence of the apsidal secular resonance.
Equations (\ref{a17-3}) and (\ref{a17-4}) are obtained with the linear secular perturbation theory,
 while to get (\ref{a17-1}) and (\ref{a17-2}),
  we use the approximations of $c_0$ and $\xi$ in (\ref{a3}) and (\ref{a3-8}).

We call the libration region defined in (\ref{a17-1}) the down-libration region, and that defined in
(\ref{a17-2}) the up-libration region.
Whether the  down-libration or up-libration is the aligned or anti-aligned resonance depends on
the sign of $(1-q\alp^{1/2})$. For $q\alp^{1/2}<1$, down-libration occurs only when $\pi/2<\dw_0<3\pi/2$ (thus it is
the anti-aligned resonance) and up-libration occurs when $\dw_0<\pi/2$, $\dw_0>3\pi/2$ (the aligned resonance).
This is the case of HD12661 system. If $q\alp^{1/2}>1$, the conclusions are reversed,
which is the case of 47UMa system.
For the critical case $q\alp^{1/2}=1$, according to (\ref{a17-1}) and (\ref{a17-2}),
all the orbits are in libration except those with $\dw_0=\pi/2, 3\pi/2$.  Fig.2 shows a typical
phase space and the evolution of an orbit in the linear secular perturbation system (1).
The jumps of $\dw$ are due to the cross of origin in the $(e_i\cos\varpi_i, e_i\sin \varpi_i)$ plane.
Fig.3 shows the resonance region in the $e_{20}-\dw_0$ plane defined by (\ref{a17-1})(\ref{a17-2})
with different $e_{10}$. The parameters $\alp$ and $q$ are taken from the planetary systems HD12661
 and 47 UMa(listed in Table 1-2). The boxes around the present configuration dots show the uncertainties of
 the elements (also listed in Table 1-2).

There are two limit cases for the criteria  (\ref{a17-1}) and (\ref{a17-2}):

(1)  $\alp\rightarrow 0$. The minimum $e_{20}$ for up-libration and maximum $e_{20}$ for down-libration
can be obtained by setting $\cos\dw=1$ in (\ref{a17-1}) and (\ref{a17-2}).
When $\alp\rightarrow 0$, the  minimum $e_{20}$ for up-libration tends to
very large and maximum $e_{20}$ for down-libration tends to zero.
So when the two planets are far away, the libration regions in $e_{20}-\dw_0$ plane
can be negligible small.

(2) $e_{i0}\rightarrow 0$.  This happens when one of the planets is in a near-circular orbit,
and it is just the case discussed in Malhotra (2002), namely
for two planets initially in nearly circular orbits, an impulse perturbation may impart
 a finite eccentricity to one planet's orbit.
When $e_{10} \rightarrow 0$,  criterion (\ref{a17-2}) is always fulfilled if
 $-\pi/2<\dw_0<\pi/2$ for $q\alp^{1/2}<1$ or $\pi/2<\dw_0<3\pi/2$ for $q\alp^{1/2}>1$.
The boundary curves are the limits of those with $e_{10}$ tends to zero in Fig.3a or Fig.3b,
with both the minimum $e_{20}$ for up-libration  and the maximum $e_{20}$ for down-libration
tend to zero. Thus  half  of the $e_{20}-\dw_0$ plane is the possible resonance region with up-libration.
So the probability of these two planets captured into apsidal resonance tends to $50\%$.
 This conclusion agrees with that of Malhotra (2002).
 It's possible that these two planets are captured in anti-aligned apsidal resonance, which depends on the sign of
   $1-q\alp^{1/2}$.
  Similarly, when $e_{20} \rightarrow 0$,  criterion (\ref{a17-1}) is always fulfilled, thus  half
  of $e_{20}-\dw_0$ plane  are possible resonance region with down-libration,
  and the probability of the two planets in apsidal resonance (either aligned or anti-aligned resonance)
  tends to $50\%$.

To compare the above  libration regions obtained by the criteria with those calculated from the secular
perturbation system,
  we integrate  the orbits of the secular perturbation system for the HD12661 system.
   Fig.4 shows the diagrams of orbits in $e_2-\dw$ plane, the initial values of the studied orbits
   have $e_{10}=0.1$, with $\dw_0=0$ for Fig.4a  and $\dw_0=\pi$ for Fig.4b.
   According to the criteria(\ref{a17-1}) and (\ref{a17-2}),
   $e_{20}>0.022$ for the aligned resonance at $\dw_0=0$,   and $e_{20}<0.375$
   for the anti-aligned resonance at $\dw_0=\pi$, which coincide with those given from  the secular perturbation system.

Fig.5 shows the contours of the excursions of $e_1$,$e_2$ for different $e_{20}$ and $\dw_0$.
As we can see, generally the orbits in apsidal resonance have relatively smaller $\Delta e_1$, $\Delta e_2$, especially
the turning points of the contour curves lie in one of libration boundary curves.
Thus from the linear secular perturbation theory, the orbits in apsidal resonance, either in aligned resonance or
anti-aligned resonance, are more stable than those in non-resonance region.

\subsection{Area of the libration region }

From  the above criteria, we can calculate the probability that the two
planets fall in apsidal resonance in the space of initial orbital elements .
We define the probability as the area of the libration region
in the $e_{20}-\dw_0$ plane for a given $e_{10}$.
 For the
down-libration case, according to (\ref{a17-1}), it's possible that the peak value of the $e_{20}-\dw_0$ curve
can be above unit for larger $e_{10}$ (as the $e_{10}=0.35$ case in Fig.3a, and
the $e_{10}=0.50$ case in Fig.3b). We set
$\dw_d$ as the half width of the down-libration region where the boundary curve reaches $e_{20}=1$.
Fig.3a and Fig.3b show $\dw_d$ for the $e_{10}=0.35$ and $e_{10}=0.50$ curves, respectively.
Define
\beq
   Q_d=\frac{5}{2}  \frac{e_{10}q\alp^{3/2}(1-\frac18 \alp^2)}{|1-q\alp^{1/2}|},
\eeq
then,
\beq
\dw_d= \left\{
\bea{ll}
 \arccos(1/Q_d),~~ &
   ~~  {\rm if}~  Q_d> 1 \\
 0 ,~~ &
   ~~ {\rm if}~ Q_d \le 1
\eea
\right.
\label{a18}
\eeq
and the area ratio of the down-libration area to the total area of $e_{20}-\dw_0$ plane is,
\beq
\bea{ll}
P_d &= \frac{1}{\pi} [\dw_d+ \int_{\dw_{a}}^{\dw_{a}+\frac{\pi}{2}-\dw_d}
 Q_d \cos \dw ~d\dw ] \\
 & =\frac{1}{\pi} [\dw_d+  Q_d (1-\sin \dw_d) ]
\eea
\label{a19}
\eeq
where the lower integration limit is the beginning point of the down-libration region in $\dw_0$-axis
($\dw_a=\frac{\pi}{2}$ for $q\alp^{1/2}<1$ and
  $\dw_a=\frac{3\pi}{2}$ for $q\alp^{1/2}>1$). Fig.3a and Fig.3b show the $\dw_a$ for the
  $e_{10}=0.35$  and $e_{10}=0.50$ curves, respectively.
As one can see, the area ratio for the down-libration  increases linearly with $e_{10}$ when $e_{10}$ is small,
since in the interval one has $\dw_d=0$ in equation(\ref{a19}). However, for larger $e_{10}$, the increase of area ratio
is no longer linear since $\dw_d \neq 0$.

  Similarly, for the up-libration resonance, if we set $\dw_u$ as the half wide of
 the up-libration region when the boundary curve meets $e_{20}=1$(see Fig3)
 and define
\beq
 Q_u=\frac25 \frac{e_{10}|1-q\alp^{1/2}|}{\alp(1-\frac18 \alp^2)},
\eeq
so,
\beq
\dw_u= \arccos(Q_u),
\label{a18-2}
\eeq
and the area ratio of the up-libration area to the total area of $e_{20}-\dw_0$ plane is,
\beq
\bea{ll}
P_u &= \frac{1}{\pi} [\dw_u-\int_{\dw_{b}}^{\dw_{b}+\dw_u}
 Q_u \frac{1}{\cos \dw} ~d\dw ] \\
 & =\frac{1}{\pi} [ \dw_u- Q_u
  \ln (\frac{1+\sin \dw_u}{\cos \dw_u}) ]
\eea
\label{a19-2}
\eeq
where the lower integration limit $\dw_b$ is the center of the up-libration region
($\dw_b=0$ for $q\alp^{1/2}<1$ and $\dw_b=\pi$ for $q\alp^{1/2}>1$, see Fig.3).

In the early evolution of planetary systems,
both $q$ and $\alp$  may vary due to planetary formation and migration.
We fix $e_{10}=0.35$ for the HD12661 systems, and see the variation of
libration area ratio with $\alp$ or $q$.
Fig.6 shows the variation of libration area ratios with $\alp$ and $q$.
In Fig.6a $q\approx 1.46 $ is fixed as the observed value. One can see
the ratios increase with $\alp$ before they reach the maximum (unit) at
$\alp_{max}=q^{-2}\approx 0.47$, which is the critical case, and then decrease.
  In Fig.6b $\alp\approx 0.32$ is fixed as the observed values.
Again the curves reach the maximum at $q=\alp^{-1/2}\approx 1.76$, the critical case.

\section{STABILITY OR ORBITS IN RESONANCE}

Since the linear secular perturbation theory is an approximation to the real three-body system, the above
criteria obtained from the linear perturbation theory has its limitation.
 To apply the linear criteria to the predicting of
the apsidal secular resonance, we integrate the orbits in a general three-body (co-planar) system,
where the longitudes of the ascending nodes and the inclinations of the two planet orbits
 are assumed to be zero ($\Omega_1=\Omega_2=0$, $i_1=i_2=0$) in the paper.
We adopt the RKF7(8) (Runge-Kutta-Fehlberg) integrator with adaptive step-sizes to integrate the orbits.
Generally the step is alternated between $0.00625$yr and $0.0125$yr, so there are 80-160 steps
 in a period of planet orbit with a semi-major axis of 1AU,
 and the final error of the  Hamiltonian of the three-body system after $10M$ years' evolution is less than
$10^{-9}$.

 We define an index to indicate whether an orbit is in libration region or not.
Choose a serial of discrete  time during the evolution of orbits (for example, every 12.5 years),
 give an index $I_n$ for each
time so that $I_n=0$ if at that time  $-\pi/2<\dw<\pi/2$ and $I_n=1$ if $\pi/2<\dw<3\pi/2$. Then the average values of
$I_n$ over very large $n$, denoted by $<I_n>$,  shows roughly the character of the orbit during the studied period of time according to
\beq
{\rm Index}= < I_n > \approx \left\{
\begin{array}{cc}
  0 & {\rm aligned~~ libration} \\
  0.5 & {\rm circulation} \\
  1 & {\rm anti-aligned~~ libration} \\
  {\rm others} & {\rm mixed }
\end{array} \right.
\eeq
Fig.7 shows the index for the HD12661 planet system  for  $10\times 51$ orbits in the interval $[0,0.5]\times [0,\pi]$
   of the $e_{20}-\dw_0$ plane(the same for the following calculations), the initial eccentricity  is
   $e_{10}=0.10$ in Fig.7a and
$e_{10}=0.35$ in Fig.7b. As one can see, most of the orbits in the libration region predicted by linear criteria
(\ref{a17-1}) and (\ref{a17-2}) are in real libration region when $e_{20}$ is small.
The discrepancies between the linear system and the three-body one
mainly occur on larger $e_{20}$ and the boundary between the libration and circulation  region.

Next we want to see whether the orbits, either in apsidal resonance or not,
 may have different stability in the general co-planar three-body model.
 The stability of an orbit in a Hamiltonian system is related with the topology (regular or chaotic)
 of the phase space,
so we calculate the largest  Lyapunov Characteristic Exponent (LCE) to indicate
whether the corresponding orbit is in a regular or chaotic region.
 The LCE at finite time $\chi(t)$ is calculated for few orbits up to $t=10$Myr (denoted
as $\chi_7$), and for most orbits up to $t=1$Myr (denoted by $\chi_6$).
Fig.8 shows the LCEs of four orbits in the four different kinds of region in the phase space.
For  curve (a), $\chi(t)$ decrease linearly with $t$, thus the orbit corresponding  to
curve (a) has zero LCE, and is in a regular region.
 Curve (b) shows a very small but non-zero LCE, so
 the orbit corresponding to (b) is in a very weak chaotic region.
 Both curves (a) and (b) have $\chi_6\sim 10^{-5}$yr$^{-1}$, $\chi_7 \sim 10^{-6}$yr$^{-1}$.
Curve (c) tends to a constant value, with $\chi_6 \approx \chi_7\sim  10^{-4}$yr$^{-1}$,
so the orbit corresponding to (c) should be in a strong chaotic region.
The orbit corresponding to curve (d) is unstable with the outer planet escape
 before $10^7$Myr, and  $\chi(t)$ for such an orbit is generally great than
$10^{-2}$yr$^{-1}$ before escape.
Thus by calculating $\chi_6$ (unit: yr$^{-1}$),  we can tell at least three different kinds of orbits:
\beq
\chi_6 \in \left\{
\begin{array}{cc}
  (10^{-5},10^{-4}]  & {\rm regular ~or ~weak ~chaos } \\
  (10^{-4},10^{-2}] & {\rm strong~~chaos} \\
  (10^{-2},10^{-1}] & {\rm unstable}
\end{array} \right.
\eeq

We calculate  $\chi_6$ for the HD12661 system with initial eccentricity
$e_{10}=0.10$ in one run and $e_{10}=0.35$ in another run, and
the other initial parameters are taken as the observed values.
Fig.9 show the results. The boundary curves for the  corresponding $e_{10}$ are also plotted in
the diagram. We find for small $e_{10}$, the orbits, whether in
aligned resonance, anti-aligned resonance or non-resonance regions, do not
show much difference about the LCEs. So in this case, whether an orbit is in apsidal
resonance do not affect much on its stability. However,  for larger initial $e_{10}$,
orbits in aligned resonance seem to be more stable since they have much lower LCE
as compared with those in the anti-aligned resonance or circulation regions with same $e_{20}$.
This example shows for larger $e_{10},e_{20}$, planets in aligned resonance regions would be relatively more stable.
From Fig.9b, we can also see the present configuration of HD12661b and HD12661c
 is in the boundary of a chaotic region,
 if $\dw_0$ is set to $130.7^0$, which is symmetric with the observed value $-130.7^0$.
 This conclusion has been obtained by Kiseleva-Eggleton  et al. (2002).

Finally, we study the role of apsidal resonance on the stability  of the orbits when the planets are migrating.
In the early stage of planet evolution, the protoplanets and
the stellar-disk might be coexisting and interacting, thus planet
migration might happen due to nebular tides(see, e.g., Ward 1997).
 We adopt the torqued three-body model as in Laughlin et al. (2002),
 and for the sake of simplicity, we consider the case that only
the outer planet experiences an azimuthal torque due to the planet-disk
interaction. We take the azimuthal
acceleration as $f_2=-2\times 10^{-6} $AU$^2$yr$^{-1}$ (which is smaller than that used in Laughlin et al. (2002),
because here we  concern the qualitative evolutions only), and study both
the forward ($t>0$) and backward ($t<0$) evolutions of orbits under
this acceleration. We calculate the orbits for the
HD12661 system with initial $e_{10}=0.35$, and $M_{10},M_{20}$ are random chosen.
All the other initial parameters of orbits are
taken from table 1.
 The evolution time span is $50000$ years.
 We find in all the studies cases(both forward and backward),
 the semi-major axis of the inner planet do not have secular changes,
 just as the results in Laughlin et al. (2002).
 Fig.10a shows the index for
the orbits during the evolution and Fig.10b shows the $a_2$ at the final time
for the forward case.
Some of the orbits which were initially in apsidal resonance region
become mixed, though, according to
Fig.6a, the libration regions are enlarged due to the
increase of $\alp$ (from 0.32 to approximately 0.41) by planet migration.
  For orbits in the aligned resonance, they
are more stable and with smaller final $a_2$, while for orbits in anti-aligned
resonance with larger $e_{20}$, or in circulation region, they tend to be more
unstable due to close encounters and thus have larger final $a_2$, and
most of them will escape soon in the following evolutions. Fig.11 shows the
backward case. We find the conclusions are more or less similar.
Though the libration regions shrink due to the decrease of $\alp$
by the planet migration (from 0.32 to approximately 0.23) in this case, the
configurations of apsidal resonance are generally preserved during
the migration. Thus  most of planet systems which are
observed in apsidal resonance now may be in resonance before the
migration begins. Again, the orbits in aligned resonance seem
to be more stable in the sense of having modest final $a_2$, which may be the
results of less close encounters between the
two planets.

\section{CONCLUSIONS AND DISCUSSIONS}

In this paper,we have studied the occurrence as well as the stability of  the apsidal
resonance.
The apsidal resonance occurs when the equations (\ref{a17-1})(\ref{a17-2}) are fulfilled.
We find the occurrence of apsidal resonance depends only
on the mass ratio $q=m_1/m_2$, semi-major axis ratio $\alp=a_1/a_2$(in secular systems, $a_1,a_2$ are constants),
  initial eccentricity ratio $e_{10}/e_{20}$ and the relative apsidal longitude $\varpi_{20}-\varpi_{10}$ of the
  two planets.
The criteria are based on the Laplace-Lagrange secular solution of linear perturbation theory.
 Based on these criteria, the ratio of
librating to non-librating orbits in the $e_{20}-\dw_0$ plane can be obtained analytically, which is given in
(\ref{a19})(\ref{a19-2}).
 We also find for two planets
on the orbits with large eccentricities, they can be in a stable configuration only when they are in
 aligned apsidal resonance.  When the planets are migrated under the planet-disk interactions,
 more than half of the studied orbits preserves the configurations of apsidal resonance.

The linear secular perturbation theory is applicable only when the
two planets are not in a lower order mean motion resonance. Since
in lower order resonances, the variations of $\varpi_1$ and
$\varpi_2$ are not guided by the secular dynamics, but the
resonance angles.  For example, for the $j:(k-j)$ resonance, if
the two resonance angles $\tha_1=j\lambda_1+(k-j)\lambda-k\var_1$,
$\tha_2=j\lambda_1+(k-j)\lambda-k\var_2$ librate around $0$ or
$\pi$, then the relative longitude of pericenter
$\dw=(\tha_1-\tha_2)/k$ must librate around either $0$ or $\pi$.
Thus we think in this case, the lower order mean motion resonance
and the apsidal resonance are not independent, and the former one
guide the dynamics. In the case that only $\tha_1$ librates, we
believe that the apsidal resonance is very difficult to occur,
since in this case $\varpi_1$ is guided by $\theta_1$, thus it can
not have similar variations with $\var_2$.

Beauge et al. (2002) 
find that, there may exist some asymmetric stationary solutions in mean motion resonance region,
where both the resonant angles and $\dw$ are constants with values different from 0 or $\pi$.
We think such kind of solutions are due to the mean motion
resonance and can only exist in the resonance regions,
since such kinds of apsidal resonance solutions with $\dw$ librates about constants with values different
 from 0 or $\pi$  can not be found in the linear secular perturbation system.

For the ten known multiple-planet systems (see Table 8 of Fischer et al. (2003)
for a list of the elements), the situation of whether apsidal resonance happens between
their planets can be classified roughly into three groups(see Table 3 for the extensions
of $\dw_0$ when
apsidal resonance would occur for the observed $q,\alp,\frac{e_{20}}{e_{10}}$):

(1) planets both in apsidal resonance and mean motion resonance.
 55 Cancri b and c (Fig.12a), GJ876 b and c , HD82943 b and c  are
in apsidal resonance since they are in mean resonances 3:1, 2:1 and 2:1 respectively.
 In fact, HD82943 b and c  can in apsidal resonances without in the mean motion resonance, while
 GJ876 b and c are near the boundary of libration according to the linear secular dynamics(Fig.13a).

(2) planets in apsidal resonances far away from lower order mean motion resonances.
HD12661 b and c, 47 UMa b and c, Ups And c and d(Fig.12b) are in
this type.  They are in apsidal resonance without the existence of any
strong mean resonances. Moreover, HD12661 b and c seems to be in anti-aligned apsidal resonance, which is in
the boundary of a chaotic region.

(3) planets not in apsidal resonance  either due to the negligible small libration region
in the $e_{20}-\dw_0$ plane or without suitable $\dw_0$.
The two planets in HD38529, HD168443, HD74156  are not in apsidal resonance,  since the libration regions
 in the $e_{20}-\dw_0$ are negligible small for the observed $q$,$\alp$ and $e_2/e_1$(Table 3).
HD37124 b and c have large aligned libration area for the present parameters,
and their eccentricities are not small, but they are not in resonance due to the present values of $\dw_0$ if
$\Omega_b=\Omega_c=0$ are assumed (Fig.13b).

  However, due to the unknown of the inclinations and the longitude
 of ascending nodes in the orbital fit from the observation data,
  it is still too early to make conclusions for some planet systems whether the planets are in apsidal
 resonances. For example, for the HD12661 b and c, they are believed to be in the anti-aligned libration now.
 This is achieved by assuming  $\Omega_1=\Omega_2=0$, and they are in the boundary of a chaotic region.
 Alternative choices of $\Omega_i$ (i=1,2) may
 change the conclusion. Especially if we chose $\Omega_1-\Omega_2\sim 180^0$, the two planets will in the
 aligned apsidal resonance , which is stable according to Fig.9b.
Similarly, HD37124 b and c could in apsidal resonance if suitable parameters $\Omega$ are observed.

Similar secular resonance may also happen due to nearly the same averaging  precessing rate of the ascending
 nodes between the two planets in
mutually inclined orbits. Since the inclination perturbations are isolated from
the eccentricity ones in the linear secular perturbation theory, we will address this problem
in a separate paper (Zhou and Sun, in preparation).

\acknowledgments

We would like to thank the anonymous referee for his valuable suggestions.
 This work is supported by the Natural Science
Foundation of China (No. 10233020), the Special Funds
for Major State Basic Research Project (G200077303), and a grant from the
Department of Education of China for Ph.D candidate training(20020284011).

\clearpage


\begin{figure}
\vspace{4cm}
 \includegraphics{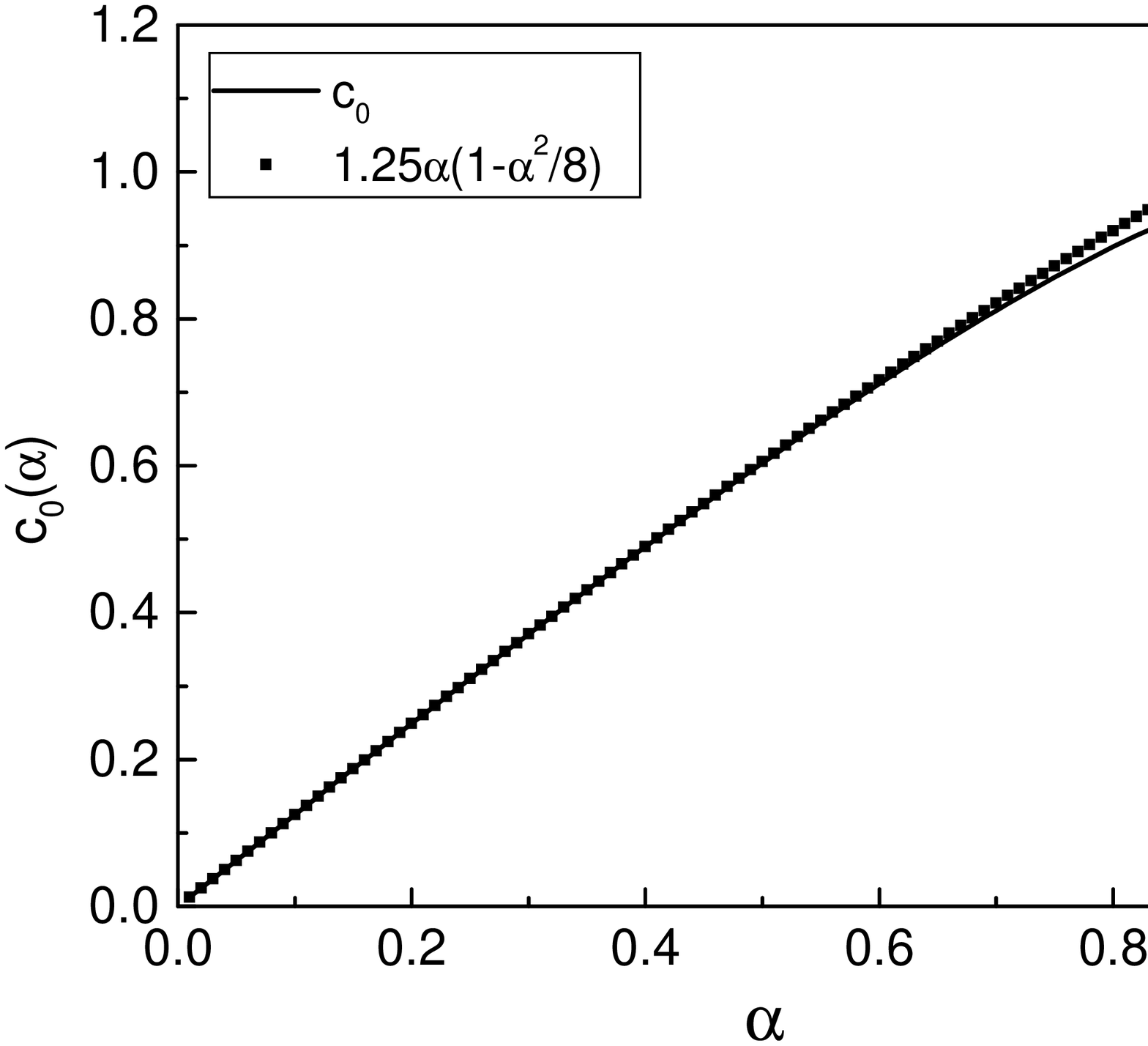}
 \caption{ The approximation of $c_0$ in equation
(\ref{a3})
 \label{fig1}}
\end{figure}

\begin{figure}
\plottwo{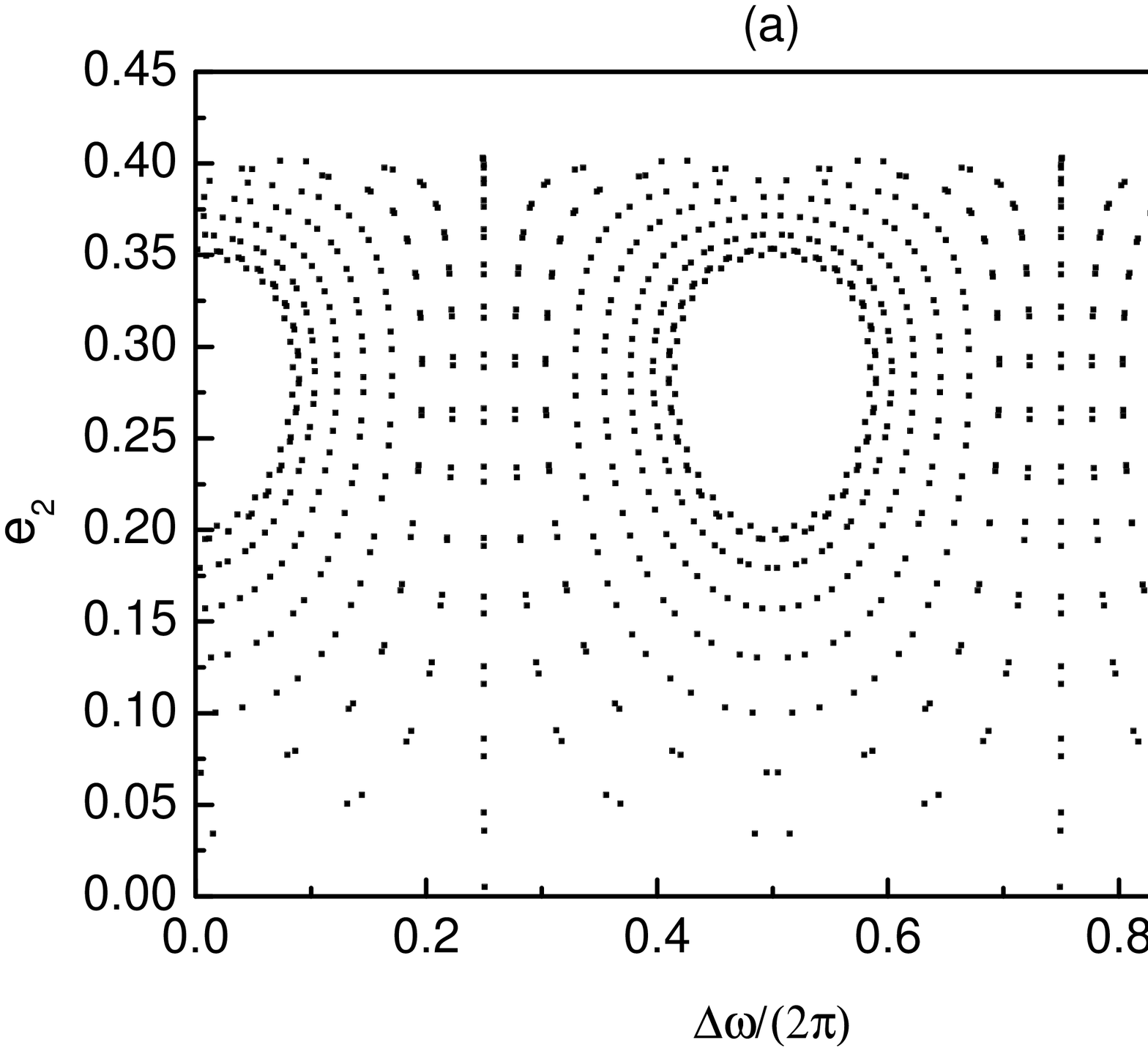}{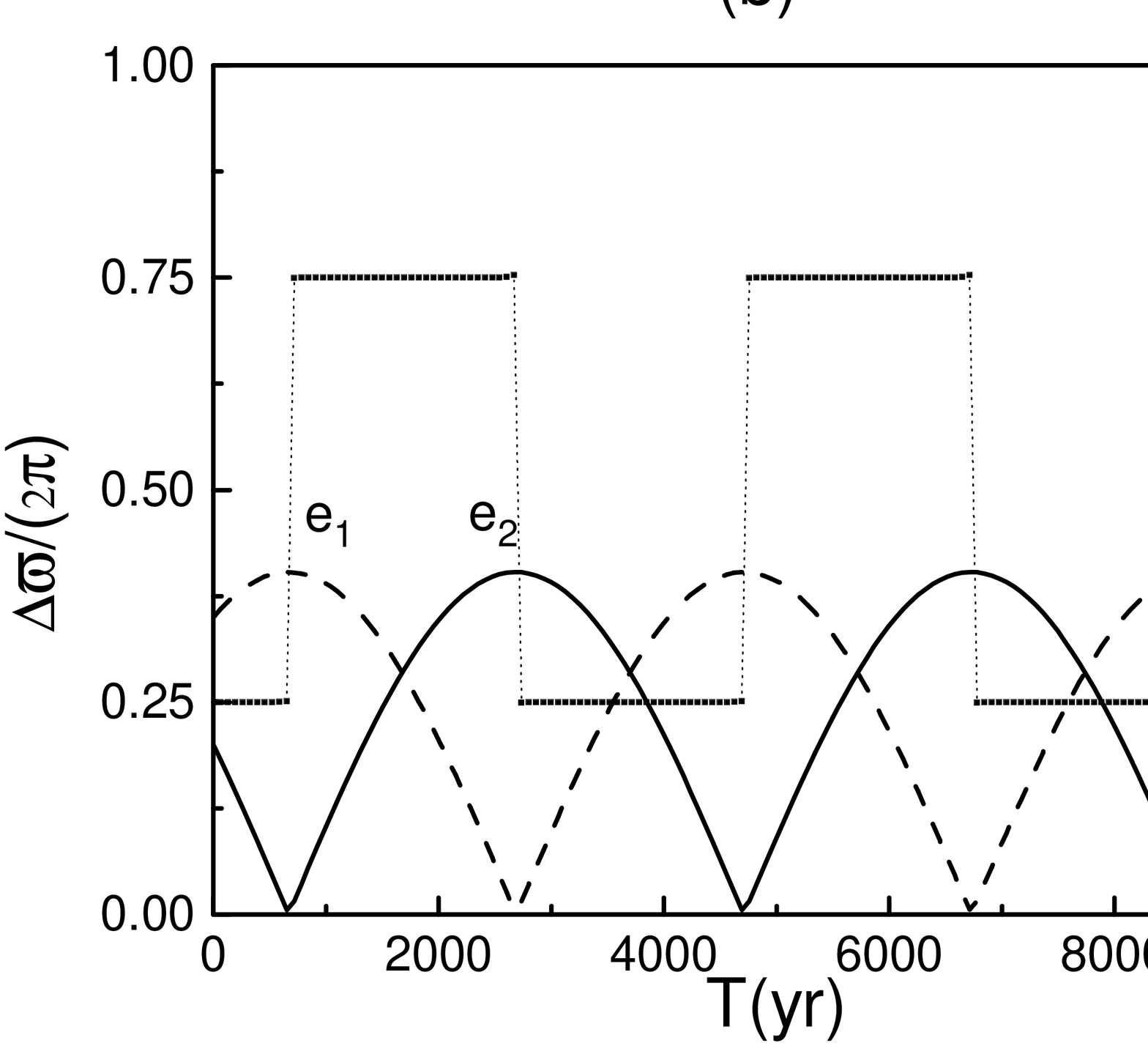}
\caption{The critical case of HD12661 with the observed $a_1$ and $q$ but with $a_2=a_1q^2\approx 1.76AU$.
(a) The phase plane $e_2-\dw$, the orbits in the diagram have the same constant $D$
as that of $e_{10}=0.35,e_{20}=0.20$. (b)Variation of $e_1,e_2,\dw$ with time for an orbit with $\dw_0=\pi/2.$
\label{fig2}}
\end{figure}
\clearpage

\begin{figure}
\plottwo{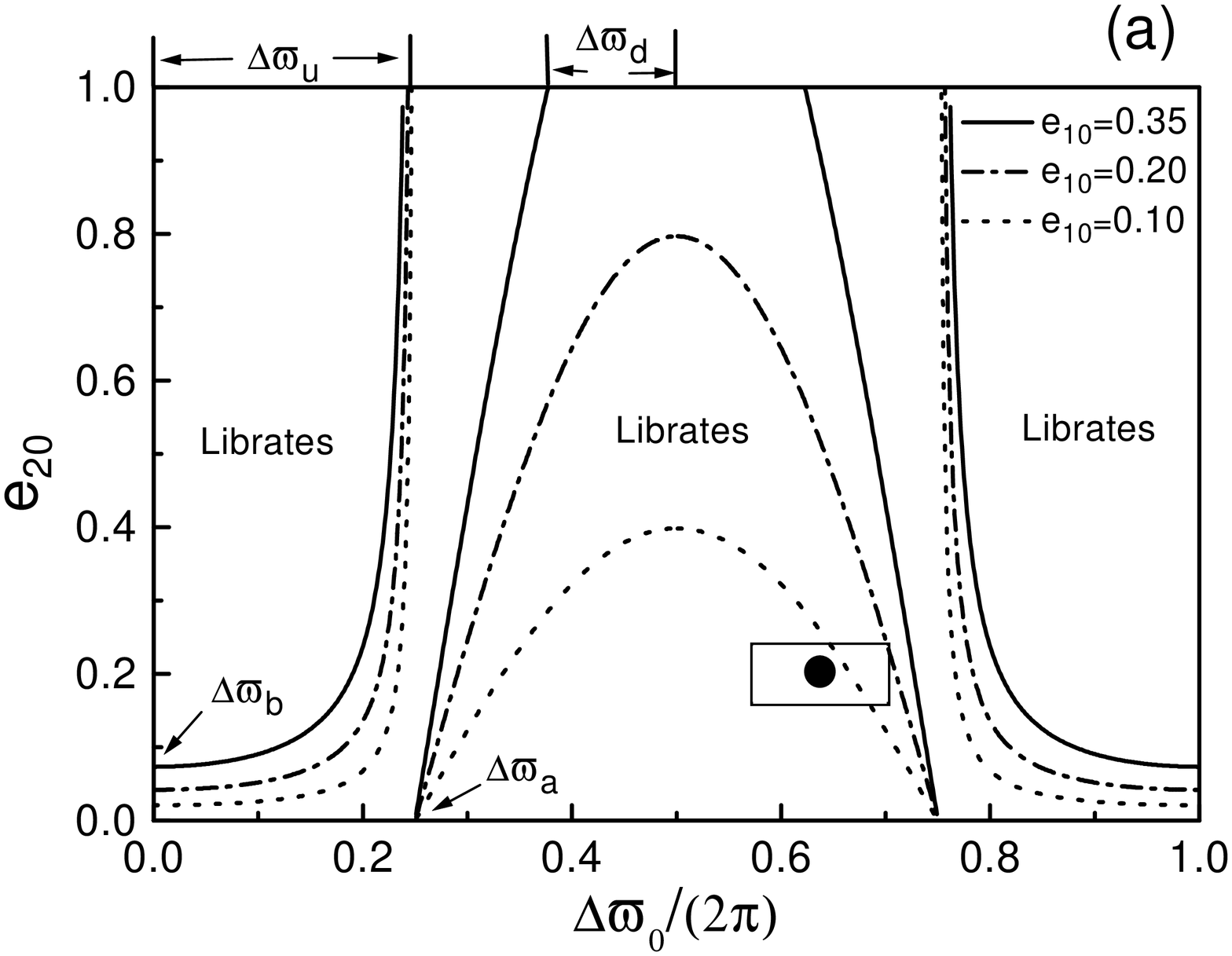}{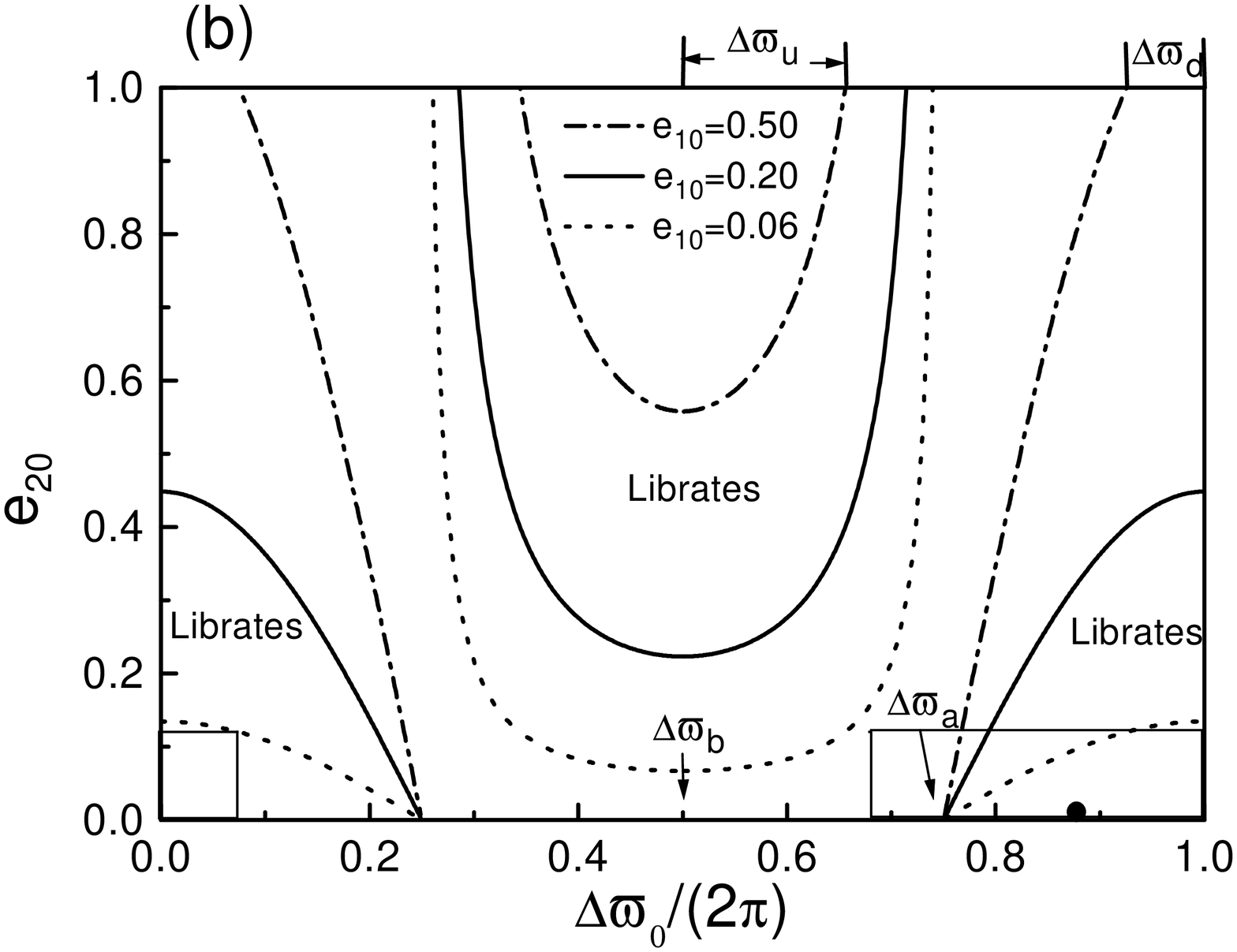}
\caption{
Libration region in $e_{20}-\dw_{0}$ plane defined by (\ref{a17-1}) and (\ref{a17-2})
 with different $e_{10}$ for $\alp$ and $q$
equal to the planetary system (a)  HD12661, and (b) 47 Uma.
 The black dots show the present configuration of the two planets in both systems, around which
 the boxes show the uncertainties of the orbital elements in Table 1 and 2.
 $\dw_a,\dw_b$ are the lower integration limits in equations (\ref{a19}) and (\ref{a19-2}),
  respectively, and $\dw_d,\dw_u$ are defined in equations (\ref{a18}) and (\ref{a18-2}),
  respectively.
 \label{fig3}}
\end{figure}

\begin{figure}
\plottwo{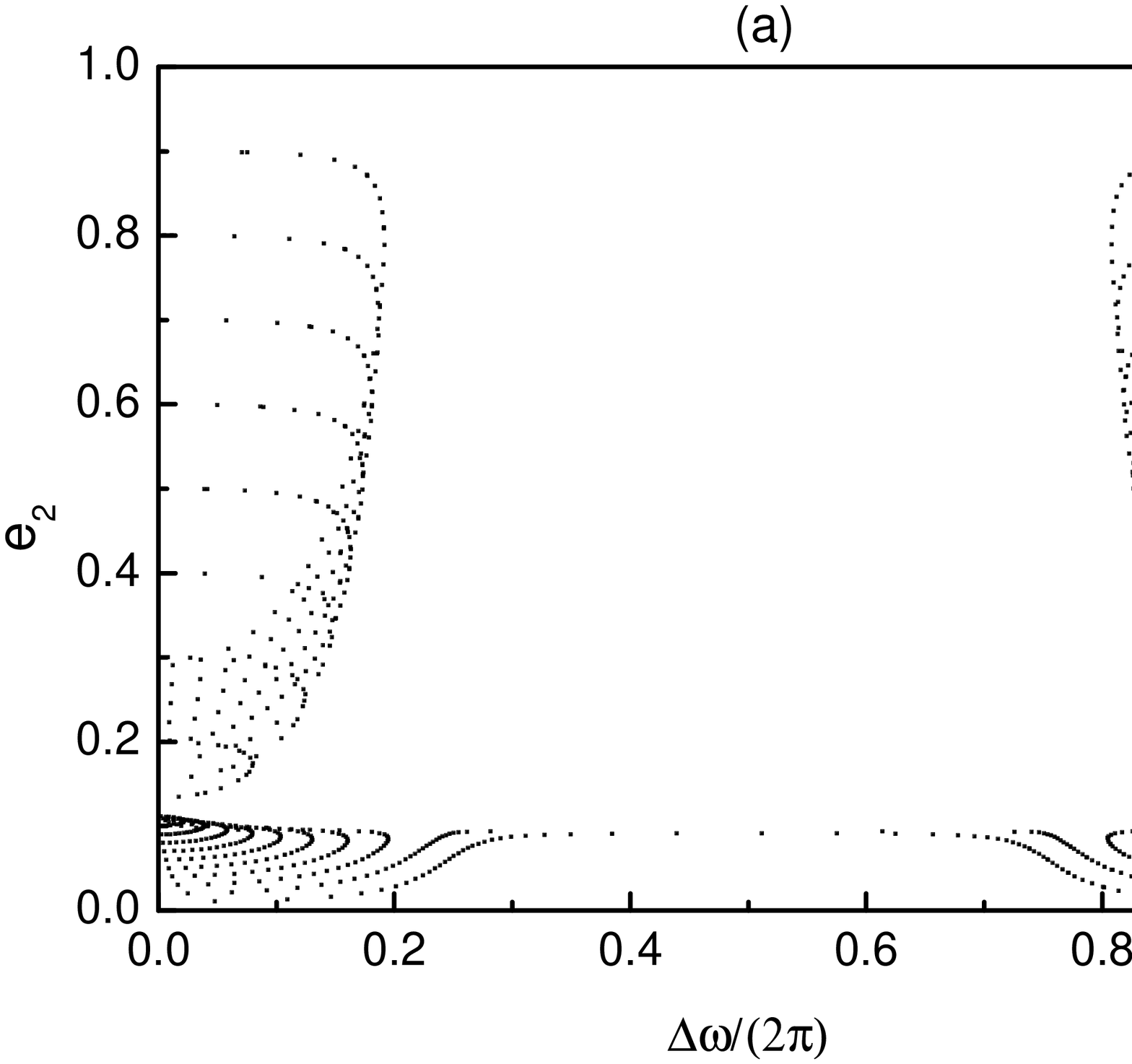}{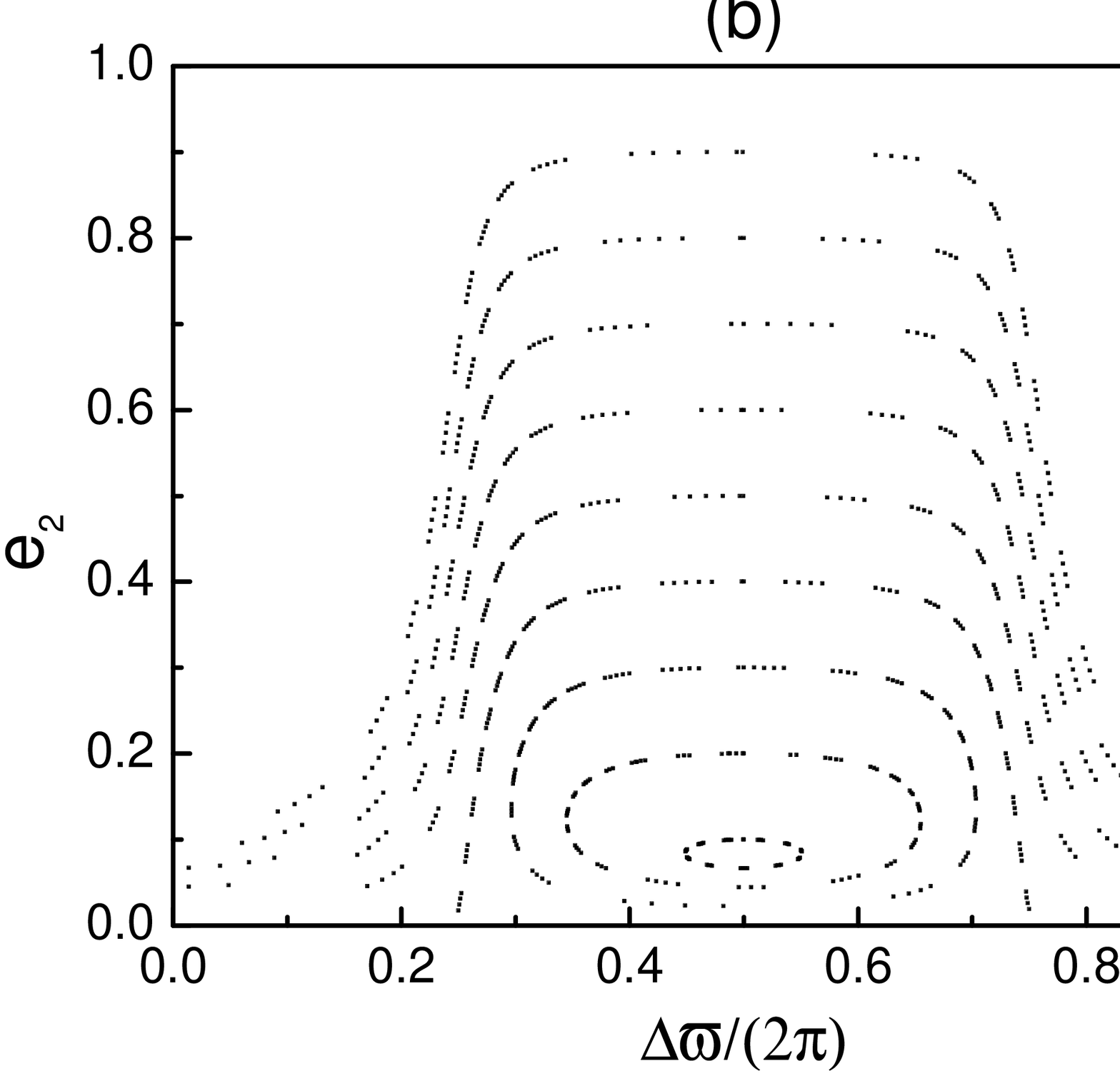}
\caption{Orbital diagrams of planet
system HD12661 with initial values (a) $e_{10}=0.1$,$\dw_0=0$;
(b)$e_{10}=0.1$,$\dw_0=\pi$ \label{fig4}}
\end{figure}

\begin{figure}
\plottwo{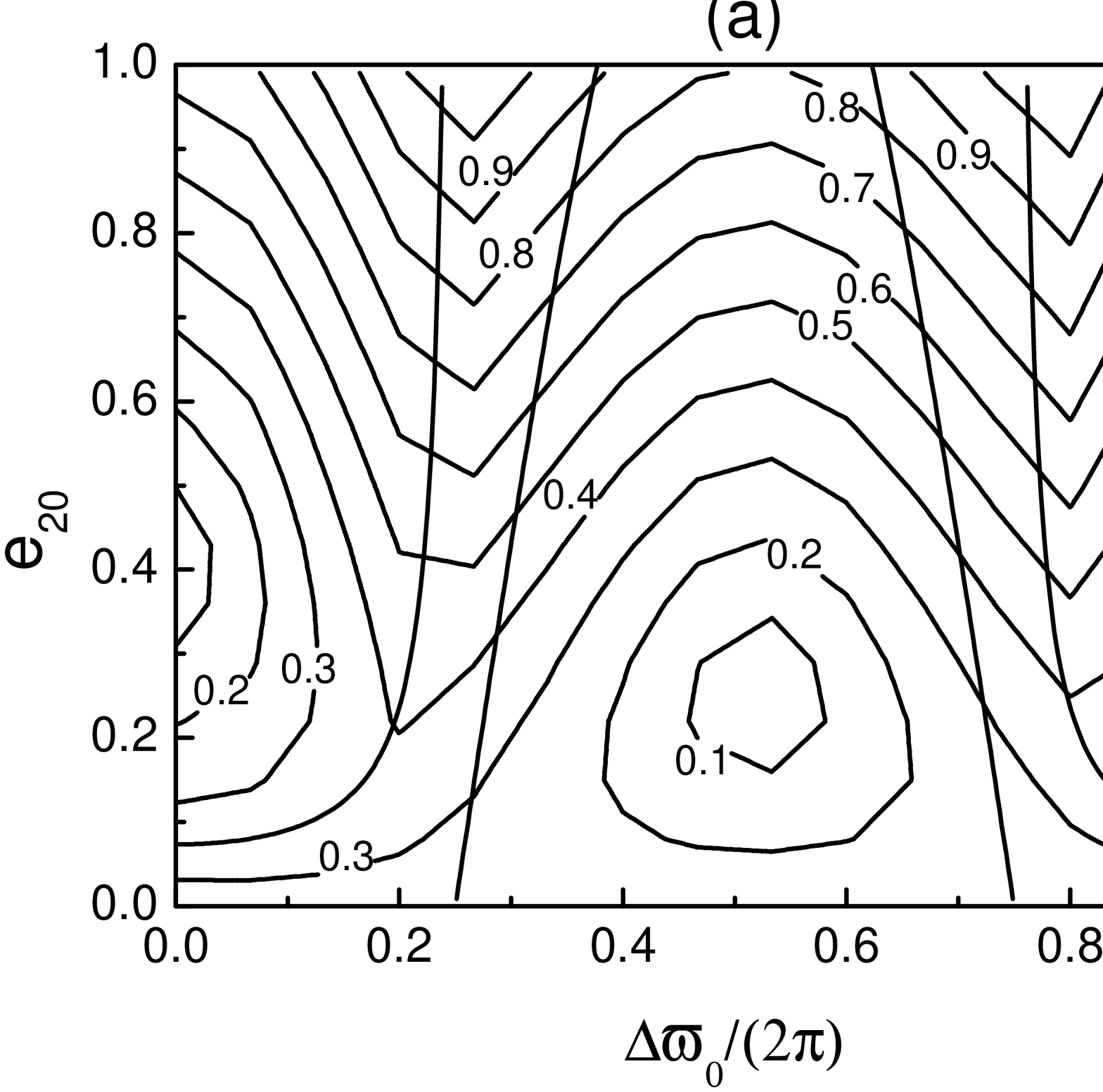}{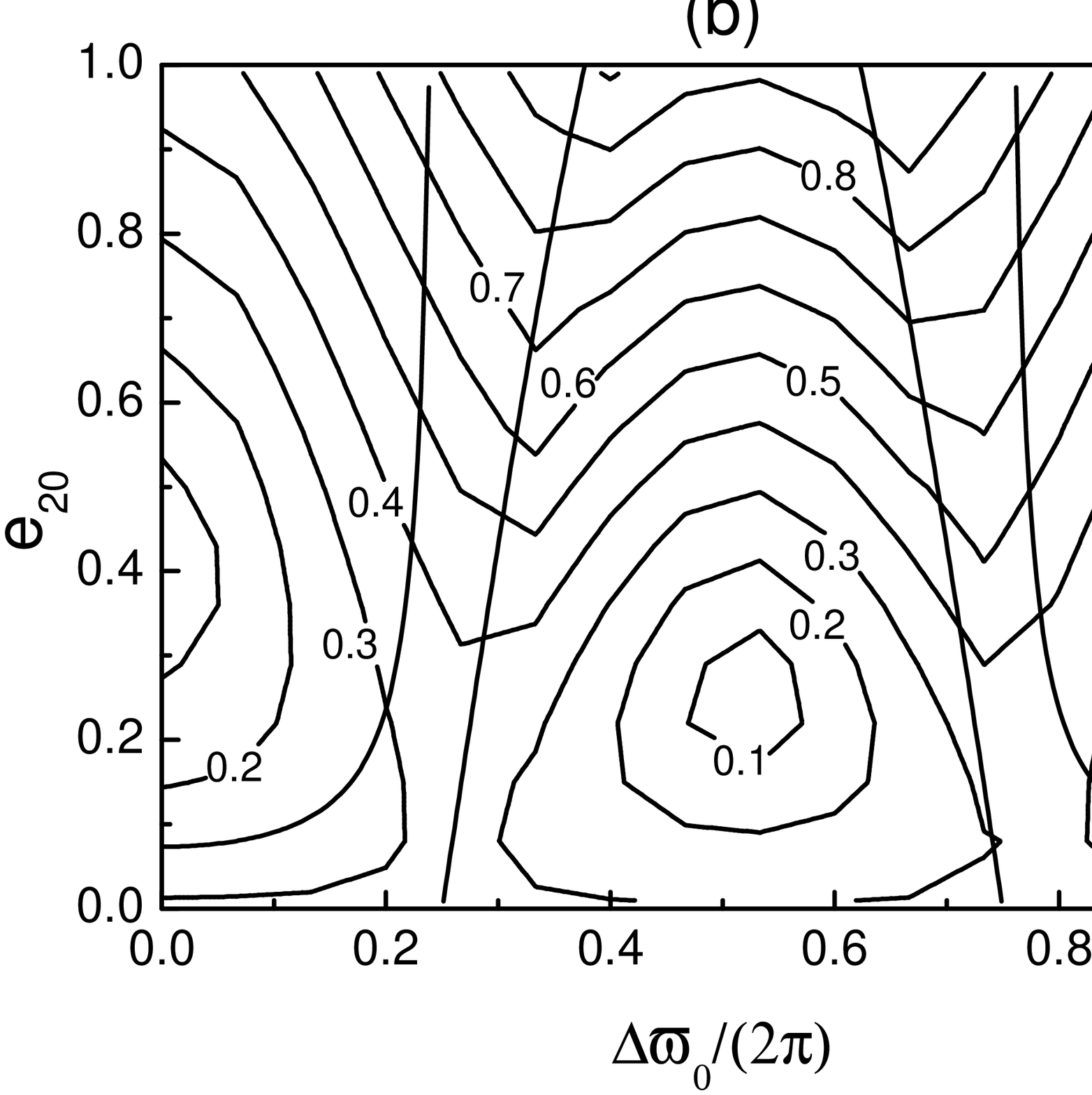}
\caption{Contours of $\Delta e_1, \Delta e_2$ defined by (\ref{de}) in the initial $e_{20}-\dw_{0}$ plane
  with  $e_{10}=0.35$.
 The dotted lines are the boundary of the libration region defined in (\ref{a17-1})(\ref{a17-2}).
\label{fig5}}
\end{figure}

\begin{figure}
\plottwo{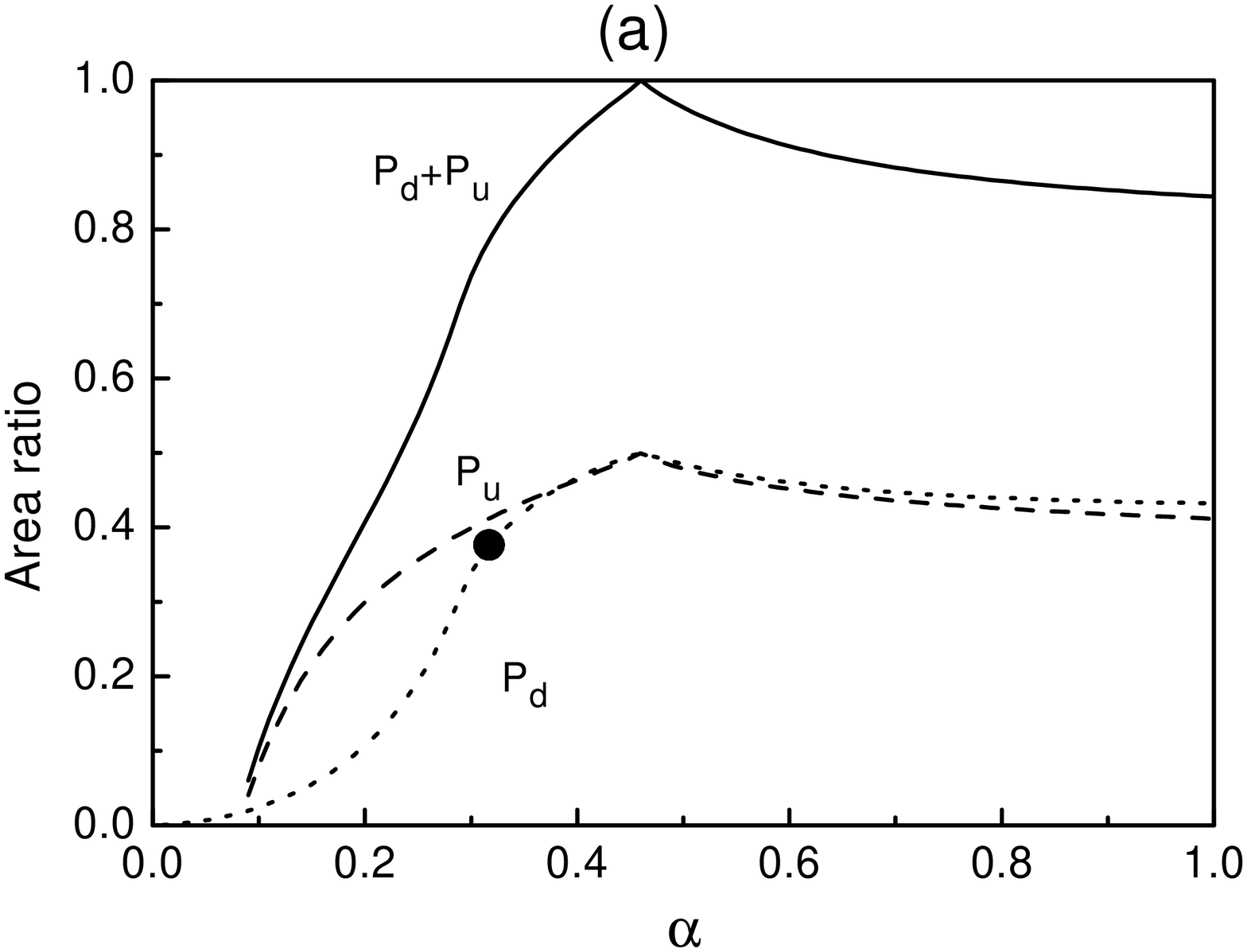}{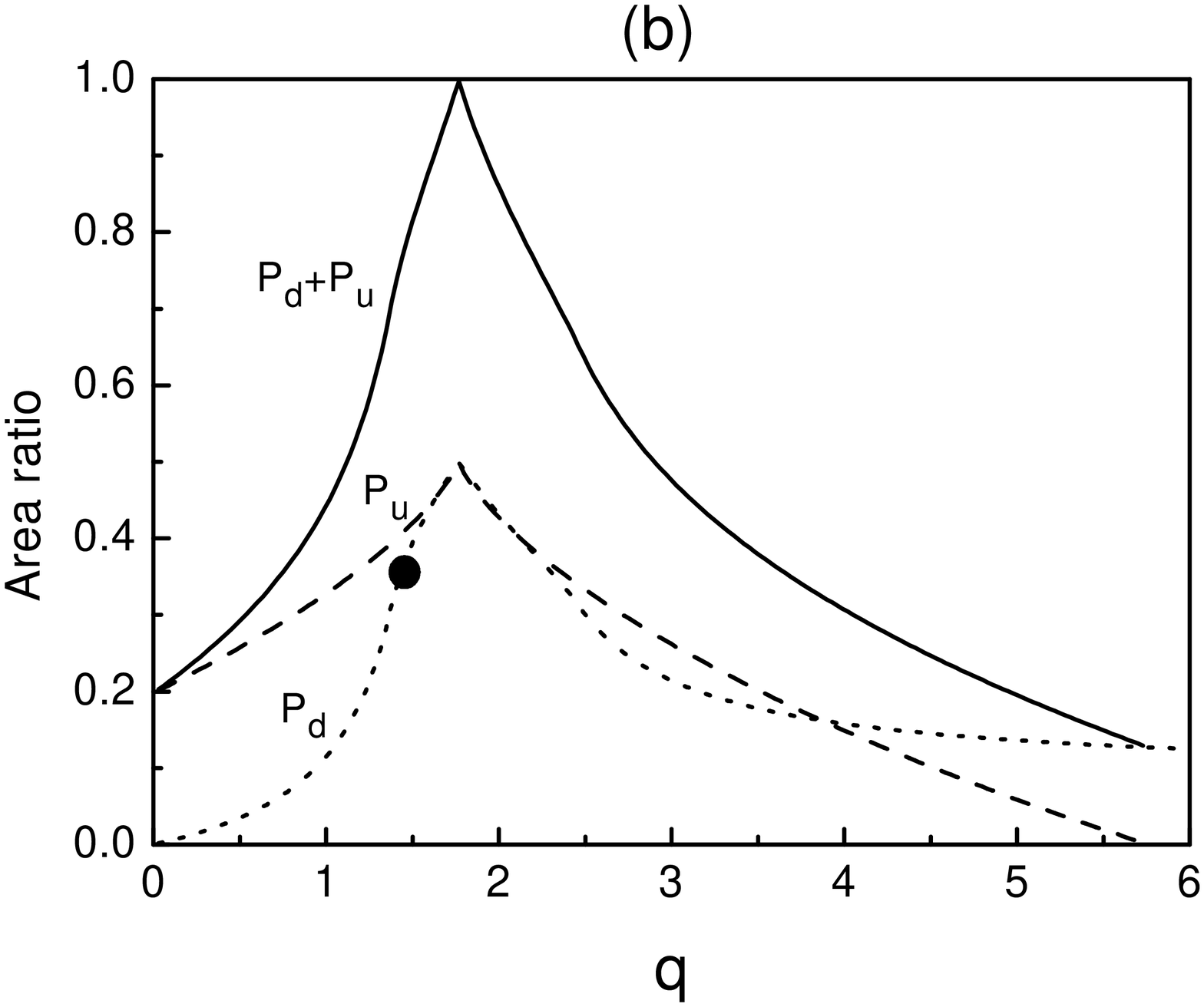} \caption{Variations of the libration
area ratios with $\alp$ and $q$ for the planet system HD12661. (a)
the observed $q\approx 1.46$ is fixed, and  (b) the observed
$\alp\approx 0.32$ fixed. In both diagrams $e_{10}=0.35$ is fixed.
 The black dot in each plot shows the location of observed configuration.
\label{fig6}}
\end{figure}

\begin{figure}
\plottwo{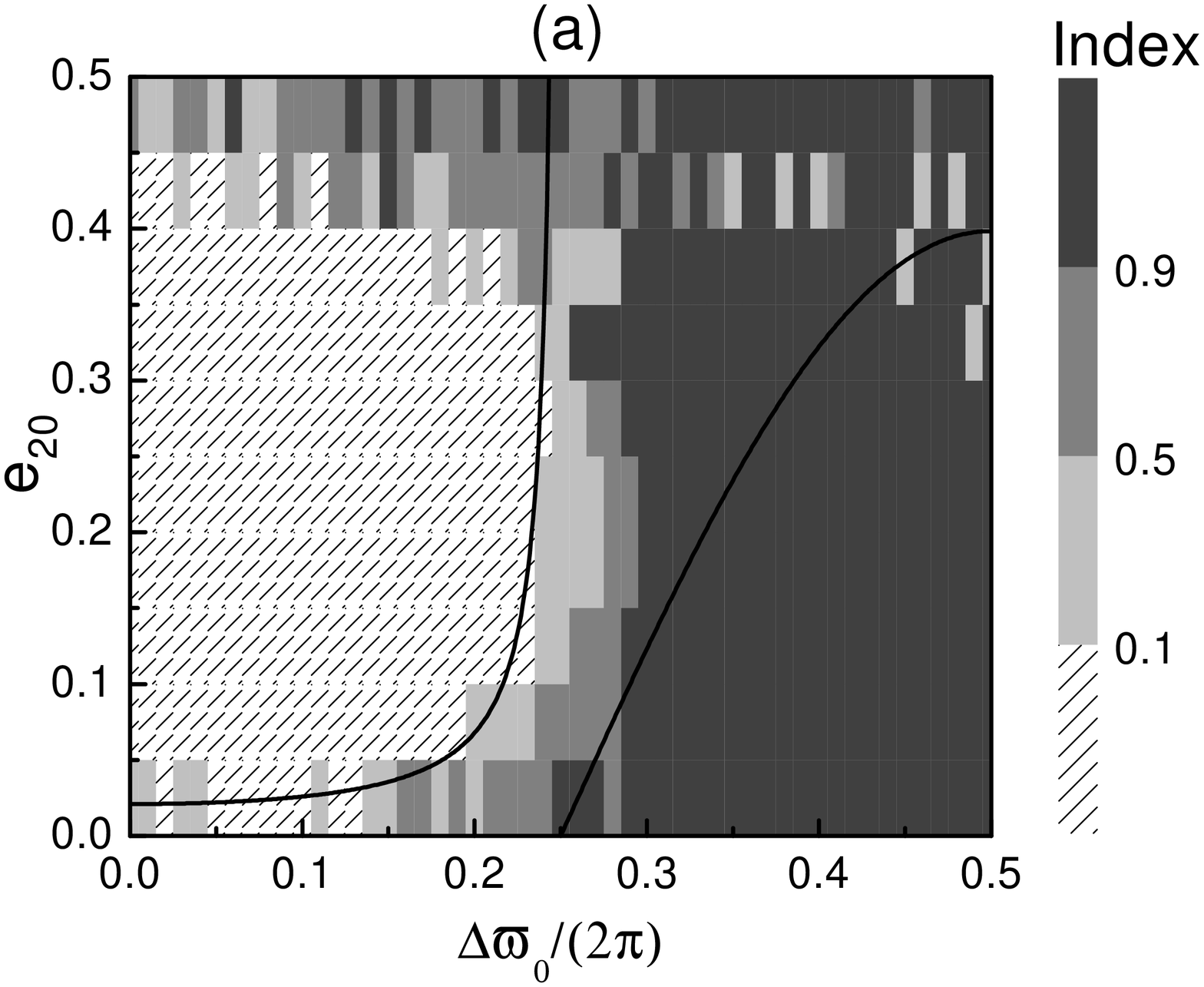}{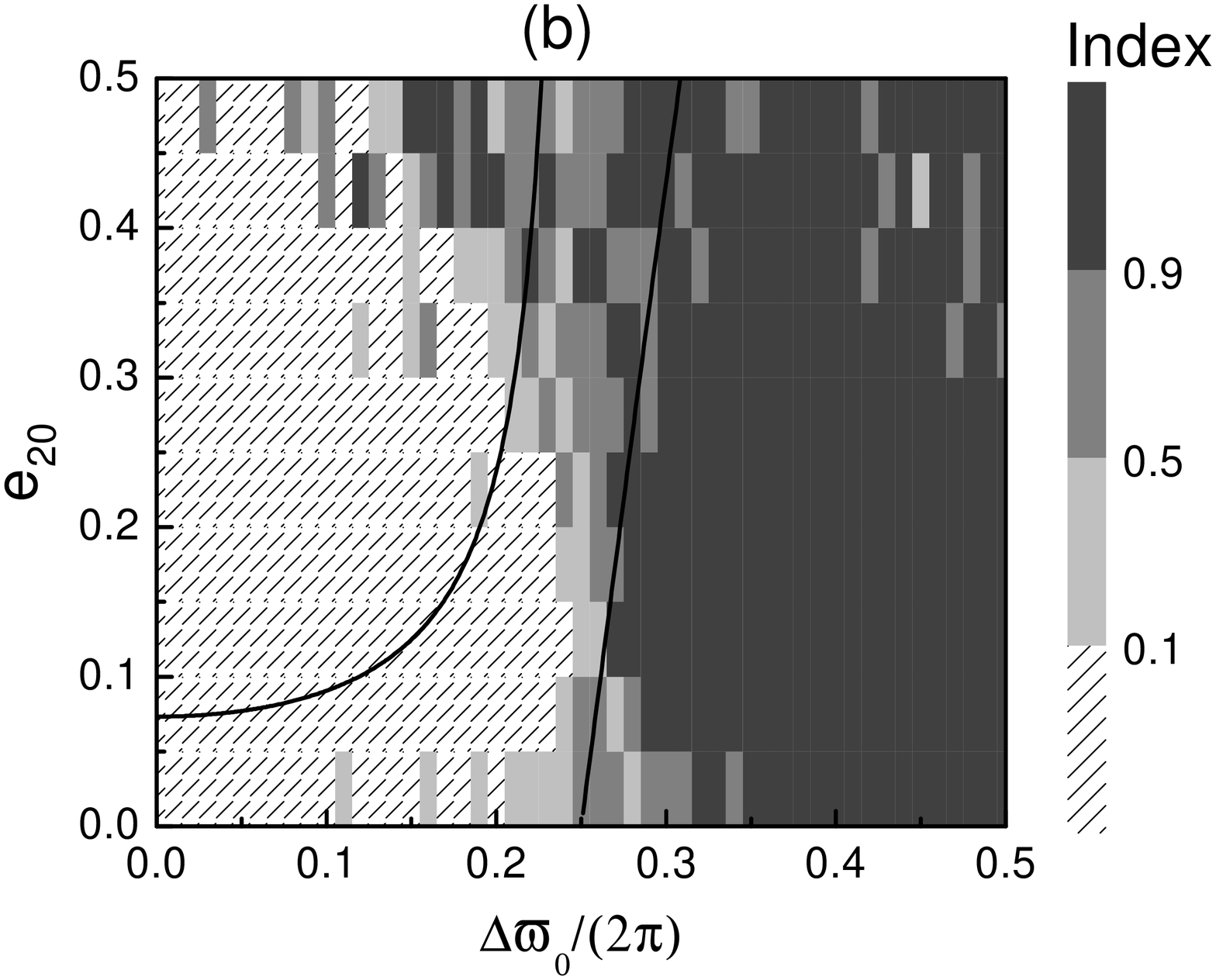}
\caption{Libration regions in the $e_{20}-\dw_{0}$ plane in the general three-body
system for the HD12661 system.
 The solid curves  show the boundary defined in (\ref{a17-1})(\ref{a17-2}).
  The initial eccentricity is (a) $e_{10}=0.10$ and (b)$ e_{10}=0.35$.
 \label{fig7}}
\end{figure}

\begin{figure}
 \vspace{4cm}
 \includegraphics{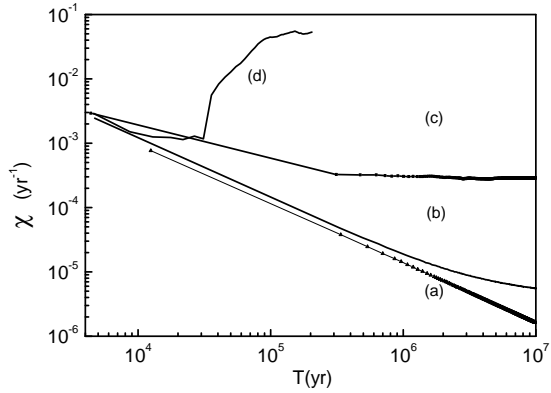}
 \caption{The LCE $\chi(t)$  for
(a) the 47Umas system with the observed orbital parameters listed
in Table 2; (b) the HD12661 system with the observed orbital parameters listed in table 1; (c)
same parameters as in (b) except $e_{20}=0.20$,~$\var_{10}=0$,~$\var_{20}=0.6\pi$;
 (d) same parameters as in (b) except $e_{20}=0.35$,~$\var_{10}=0$,~$\var_{20}=0.7\pi$, the outer planet is
 escape at time $t\approx 2.1\times 10^5$yr.
 \label{fig8}}
\end{figure}

\begin{figure}
\plottwo{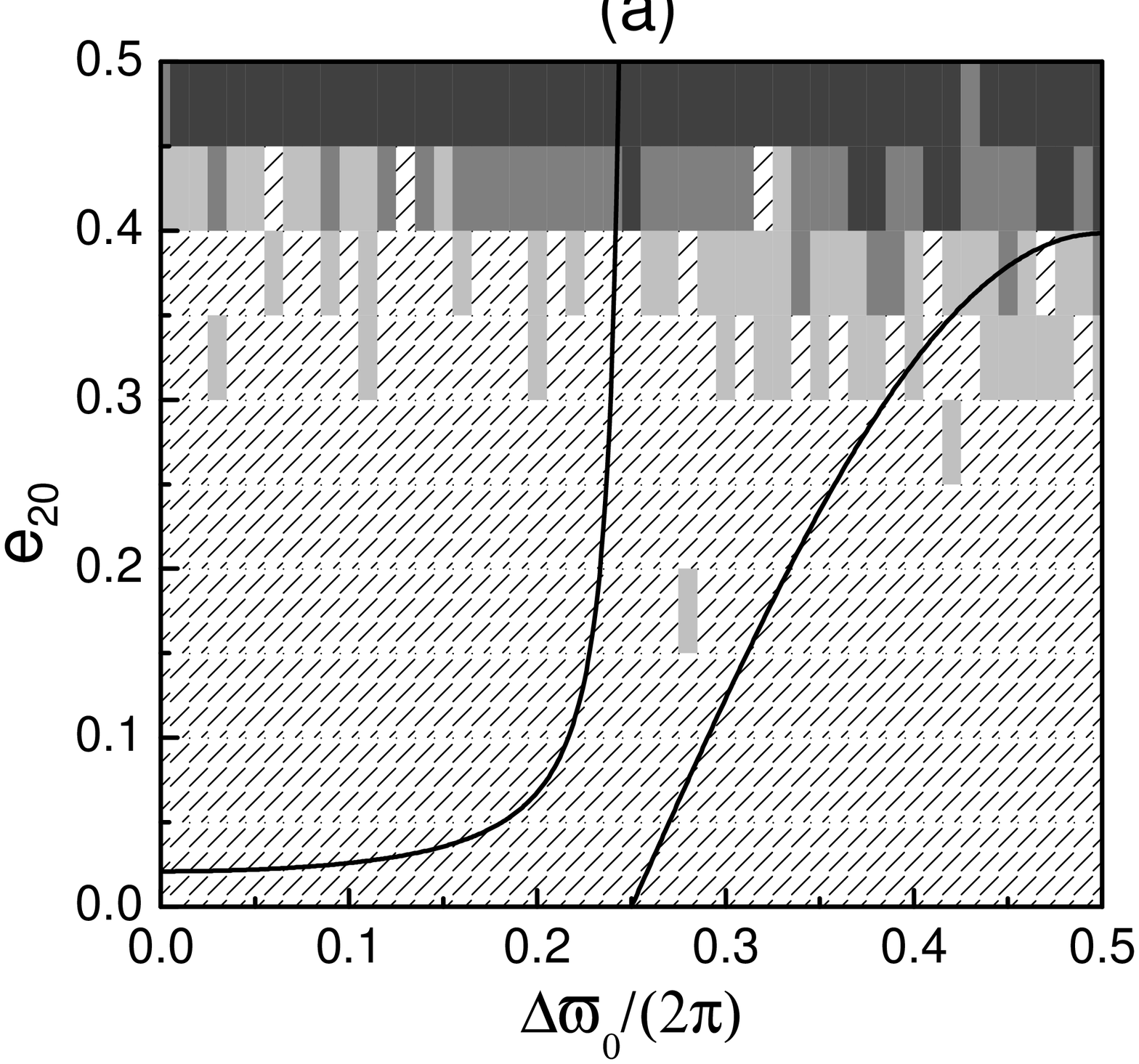}{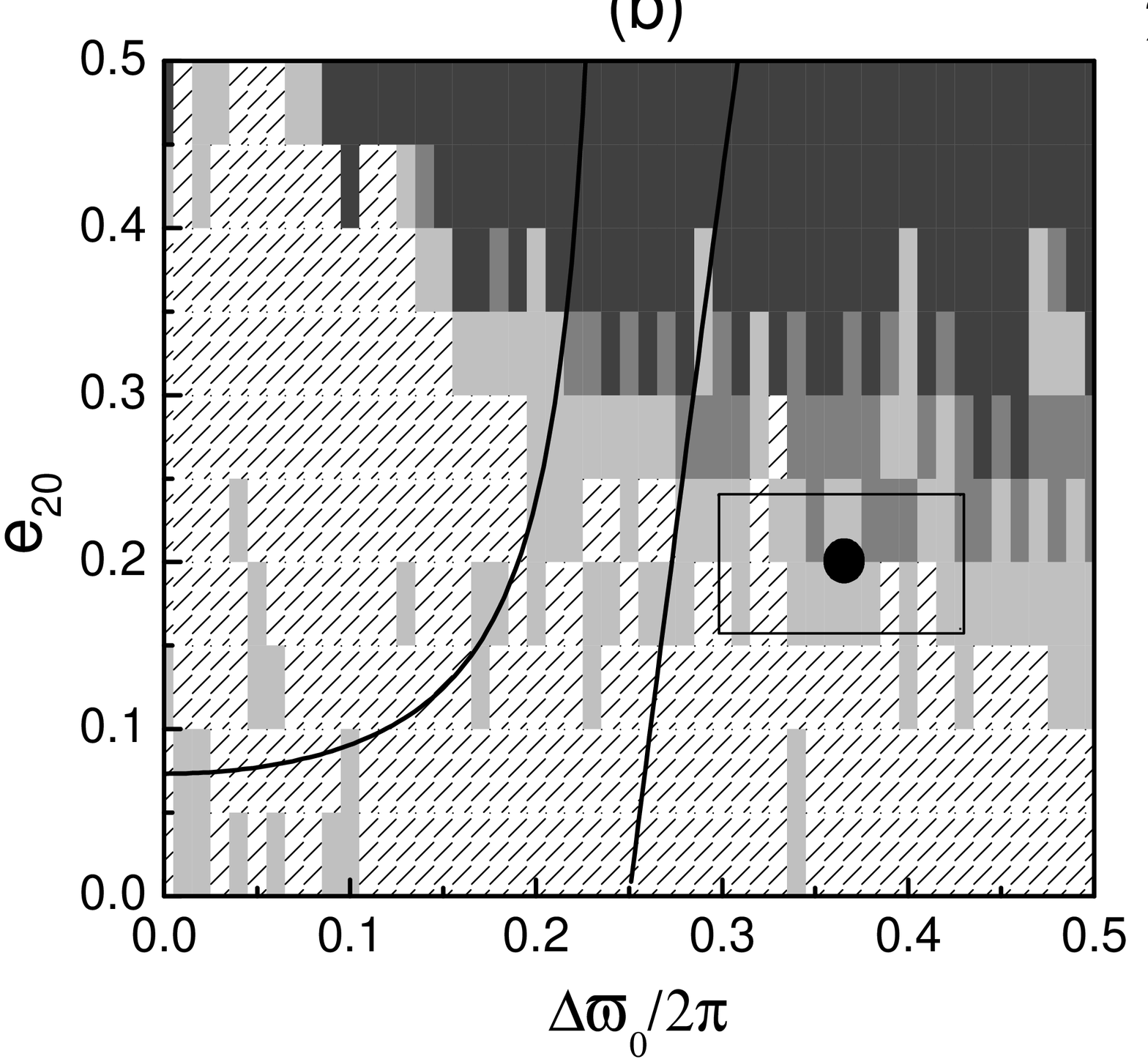}
\caption{The LCE $\chi_6$ for HD12661 system. The initial eccentricity is(a) $e_{10}=0.10$, (b)$ e_{10}=0.35$.
The solid curves  show the boundary defined in (\ref{a17-1})(\ref{a17-2}).
The black dot in (b) shows the location of observed configuration with $\dw_0=130.7^o$, around which
 the box show the uncertainties of the orbital elements.
 \label{fig9}}
\end{figure}

\begin{figure}
\plottwo{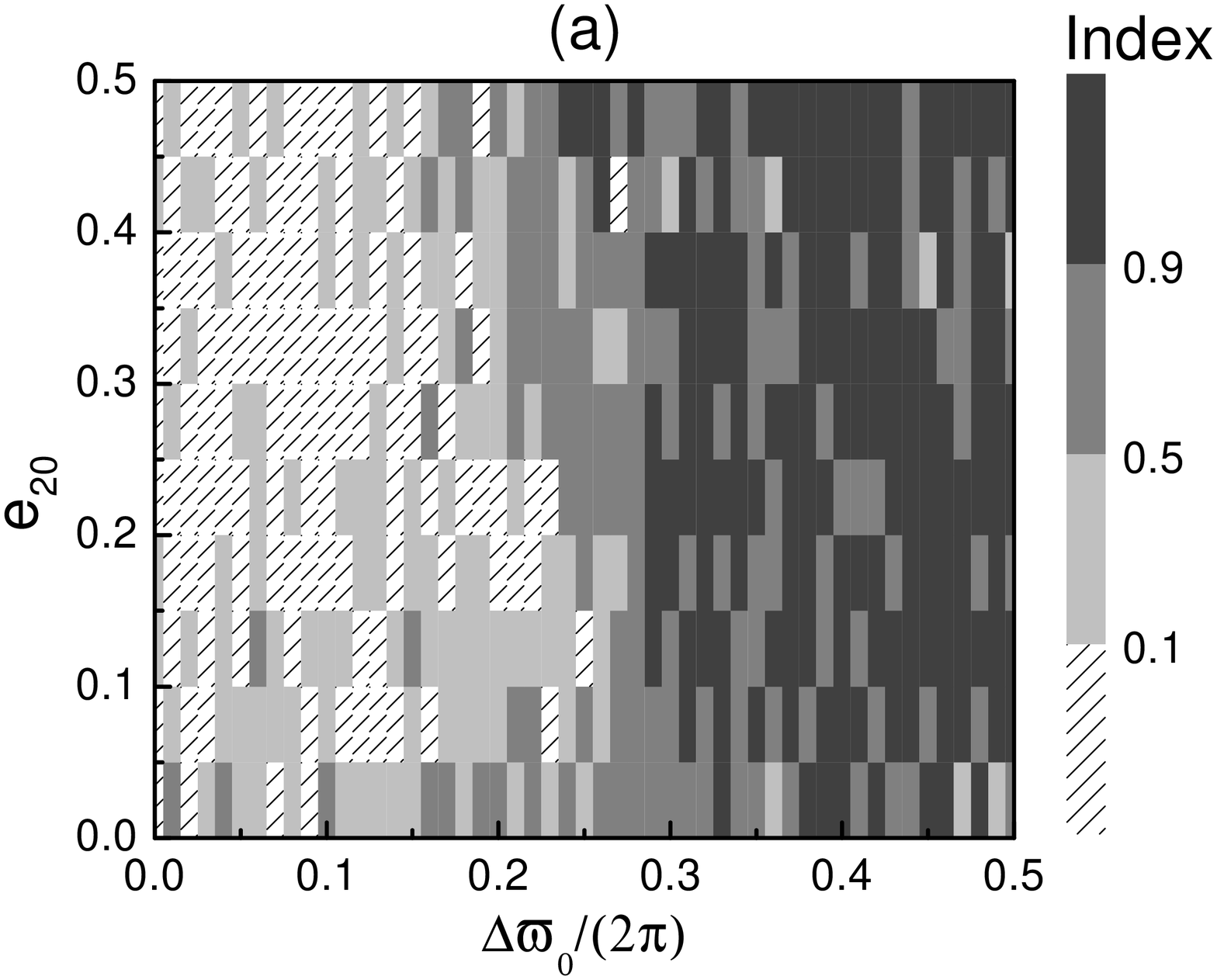}{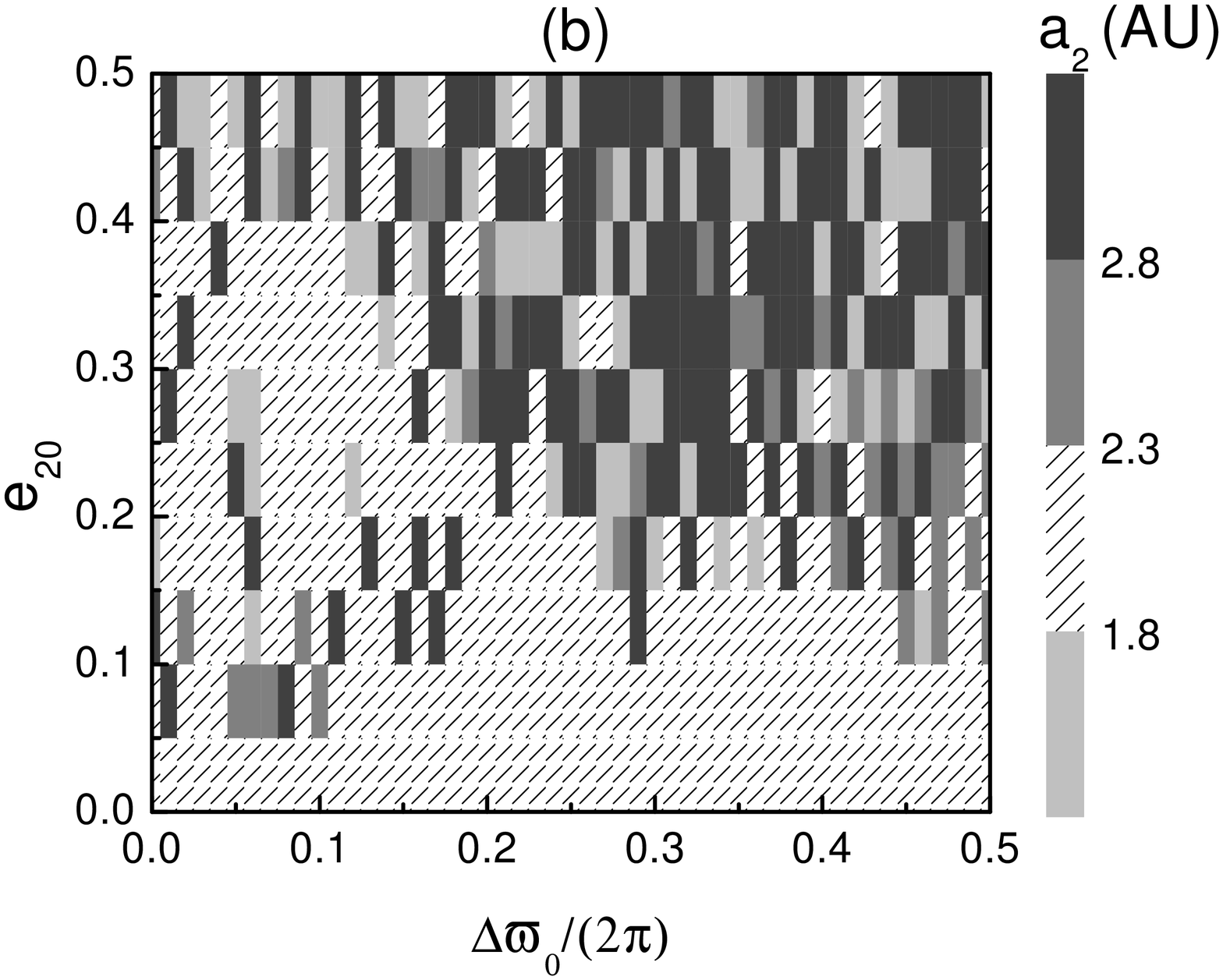}
\caption{(a) The index of libration region and (b) the relative semi-major axis of two planets
HD12661 systems under azimuthal acceleration $f_2=-2\times 10^{-6} {\rm AU}^2 {\rm yr}^{-1}$ .
The initial eccentricity is $e_{10}=0.35$ and $M_{10},M_{20}$ are random chosen. The time evolution is $t=50000$yr.
\label{fig10}}
\end{figure}


\begin{figure}
\plottwo{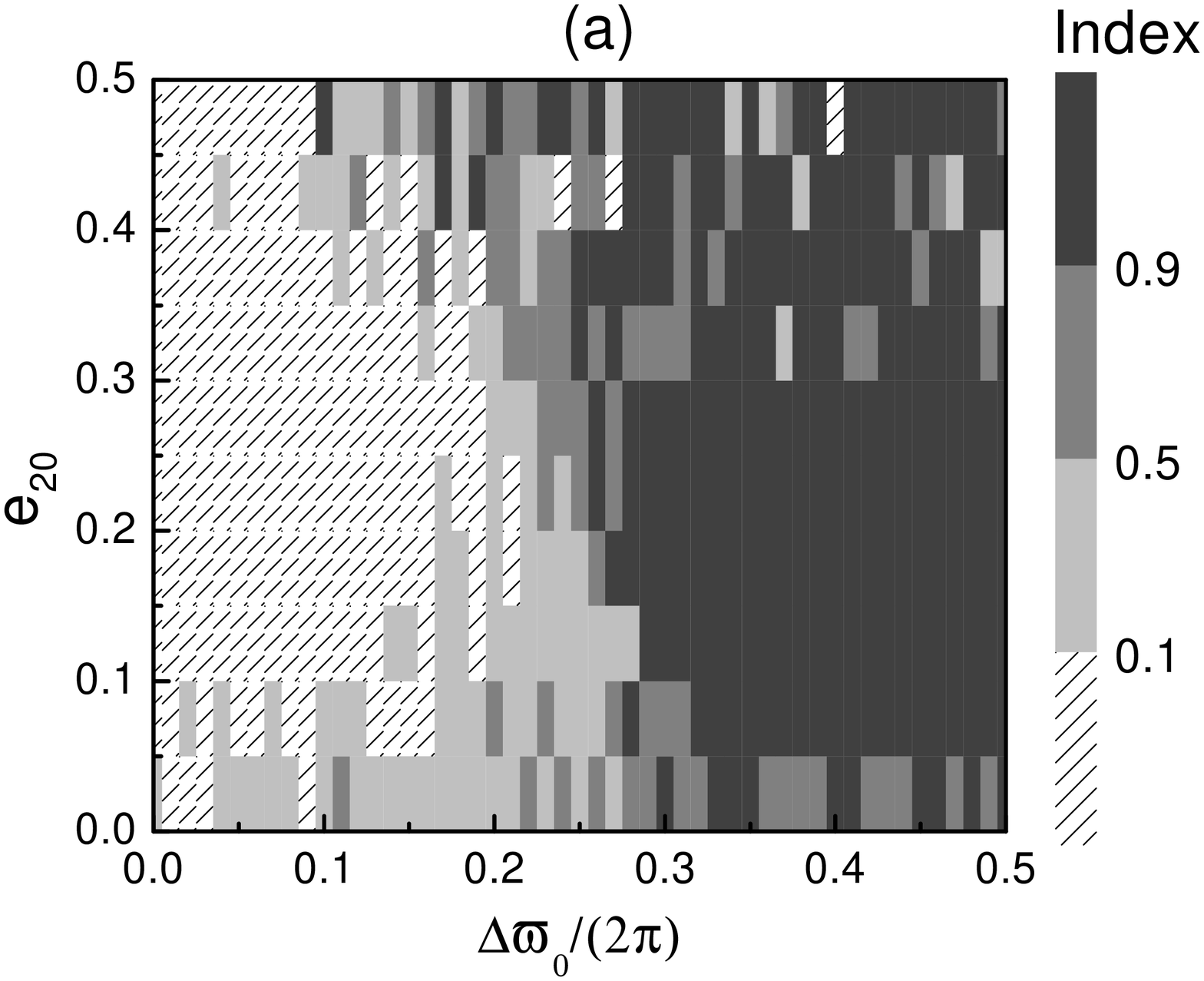}{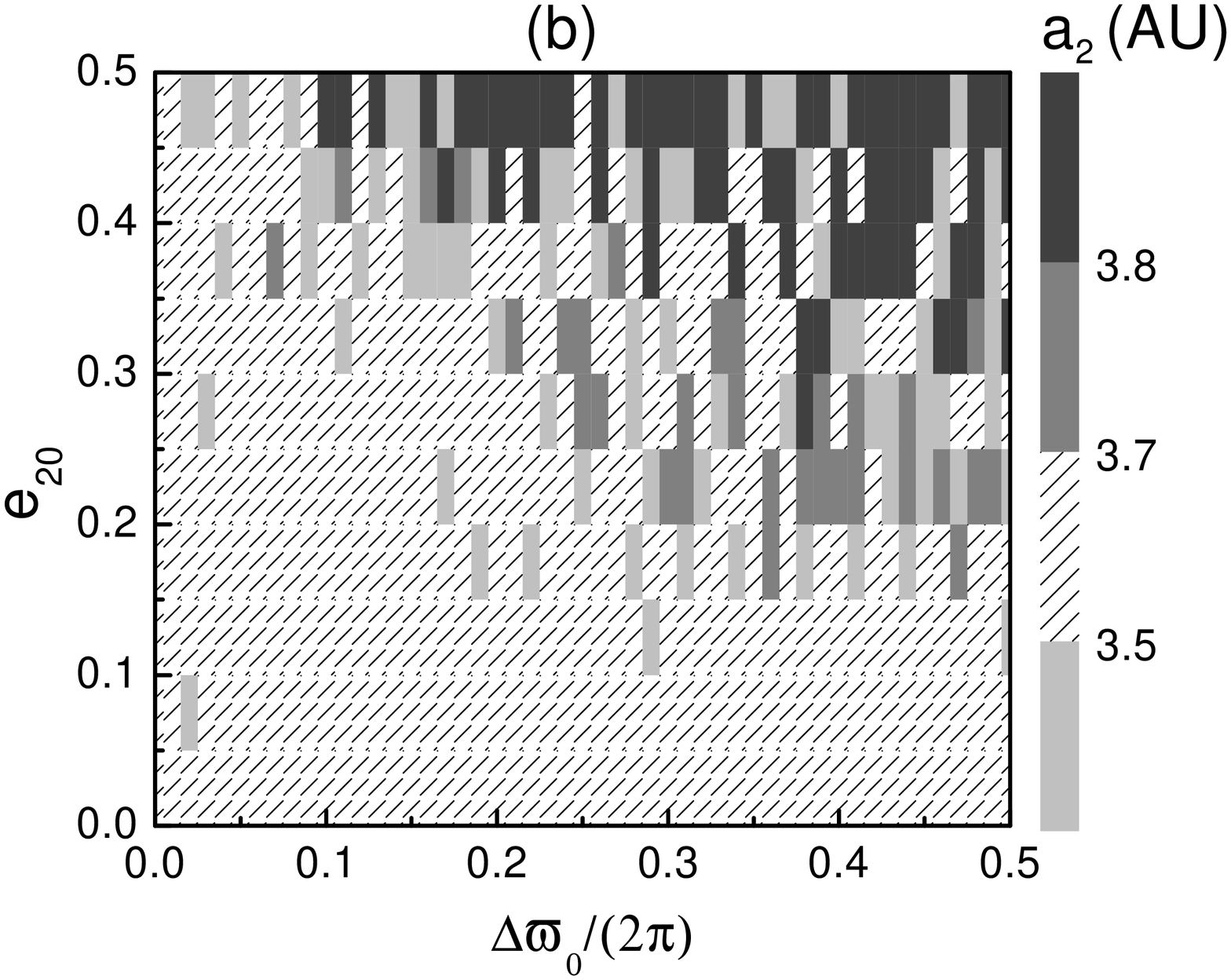}
\caption{(a) The index of libration region and (b) the relative semi-major axis of two planets
HD12661 systems under azimuthal acceleration $f_2=-2\times 10^{-6} {\rm AU}^2 {\rm yr}^{-1}$.
 The initial eccentricity is $e_{10}=0.35$ and $M_{10},M_{20}$ are random chosen.
  The evolution time is $t=-50000$yr.
\label{fig11}}
\end{figure}

\begin{figure}
\plottwo{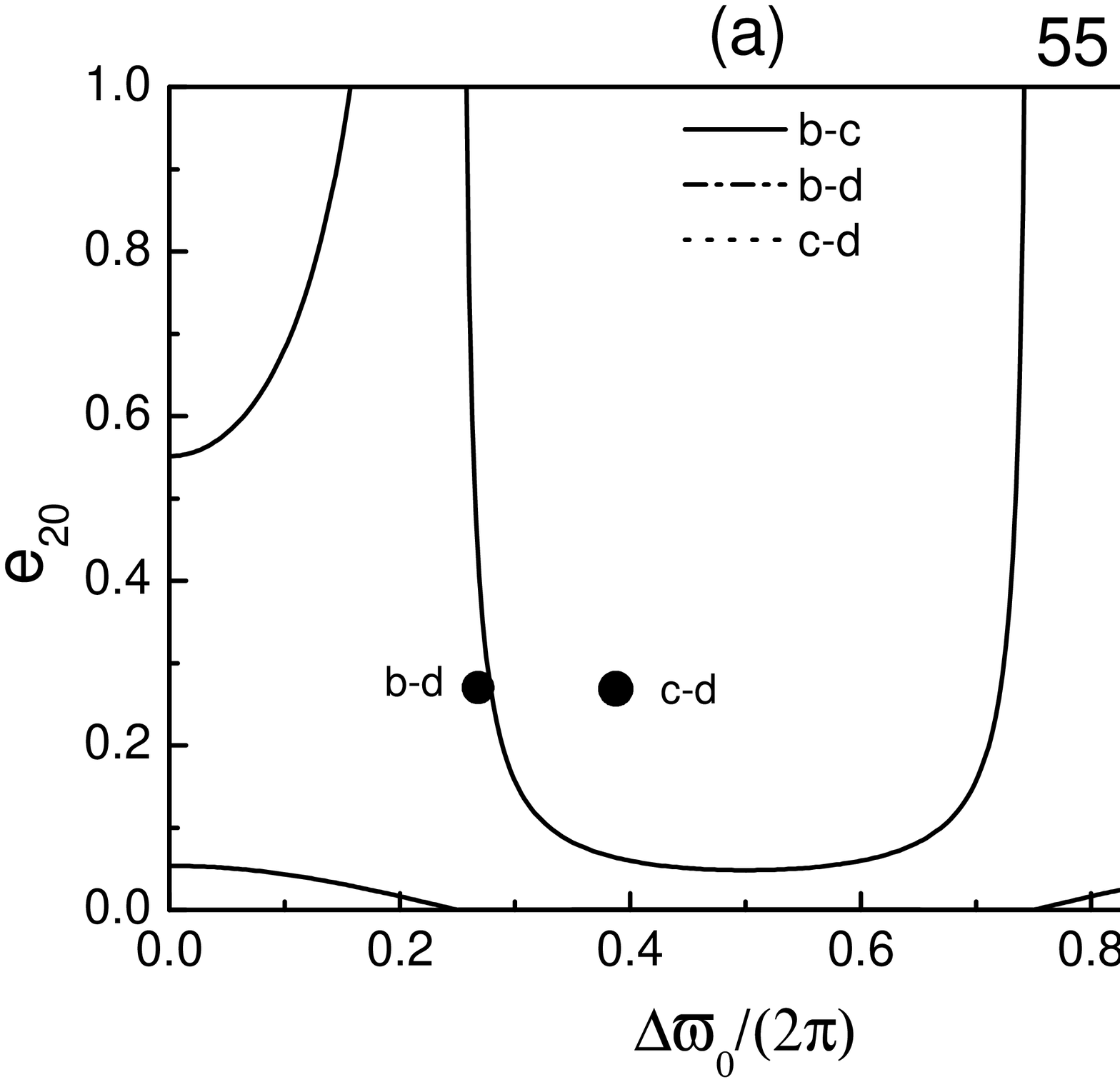}{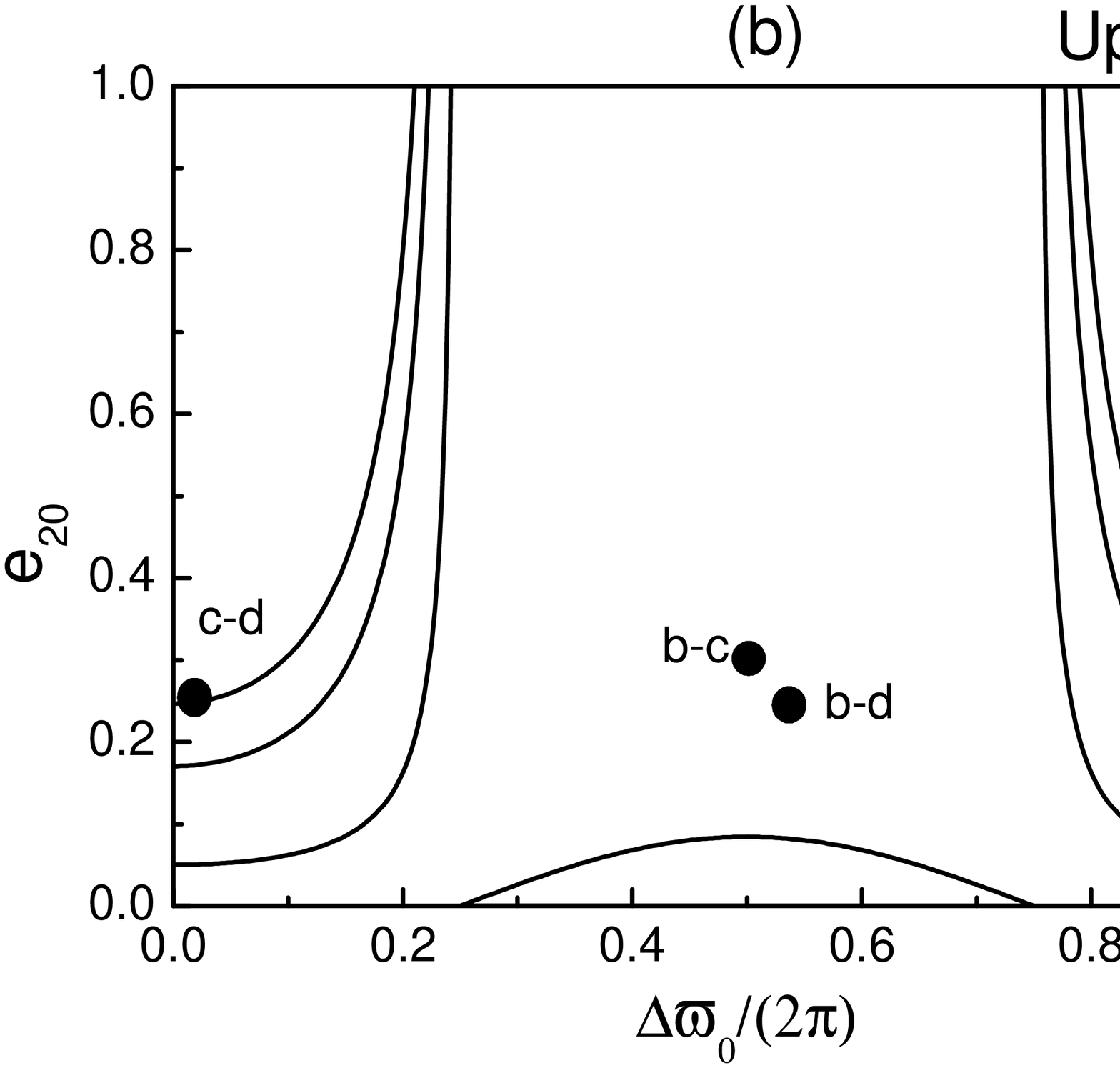}
\caption{
Libration region in the initial $e_{20}-\dw_{0}$ plane defined by (\ref{a17-1}) and (\ref{a17-2}) for
the two triple-planet systems
(a)55 Cancri ($\varpi_b=99^0$,$\varpi_c=61^0$,$\varpi_d=201^0$);
(b) Upsilon Andromedae ($\varpi_b=73^0$,$\varpi_c=250^0$,$\varpi_d=260^0$).
Invisible boundary curves are out of the range of $e_{20}$.
 The black dots show the present configuration of the two planets.
 Other orbital elements  are taken from Fischer et al. (2003).
 \label{fig12}}
\end{figure}
\clearpage

\begin{figure}
\plottwo{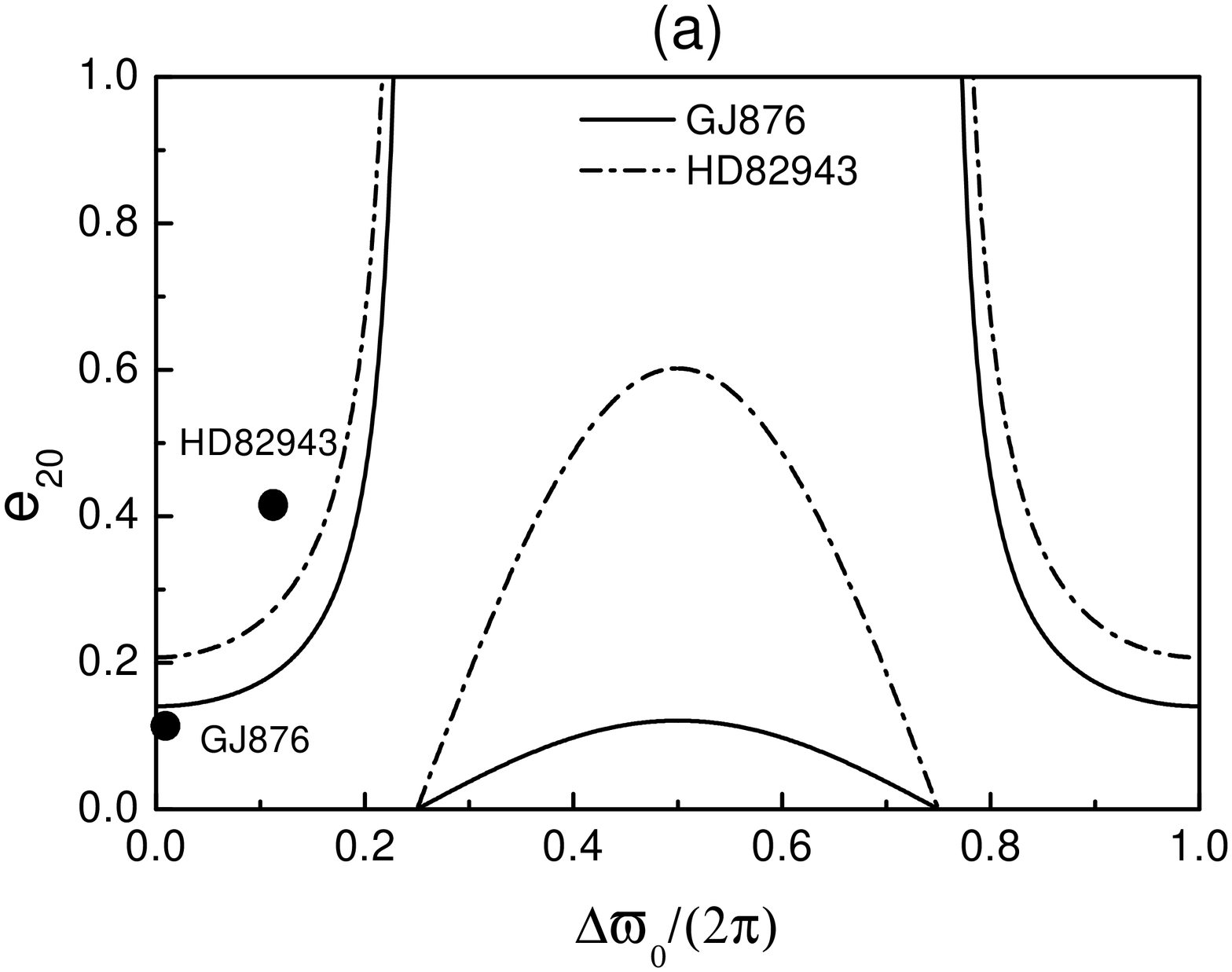}{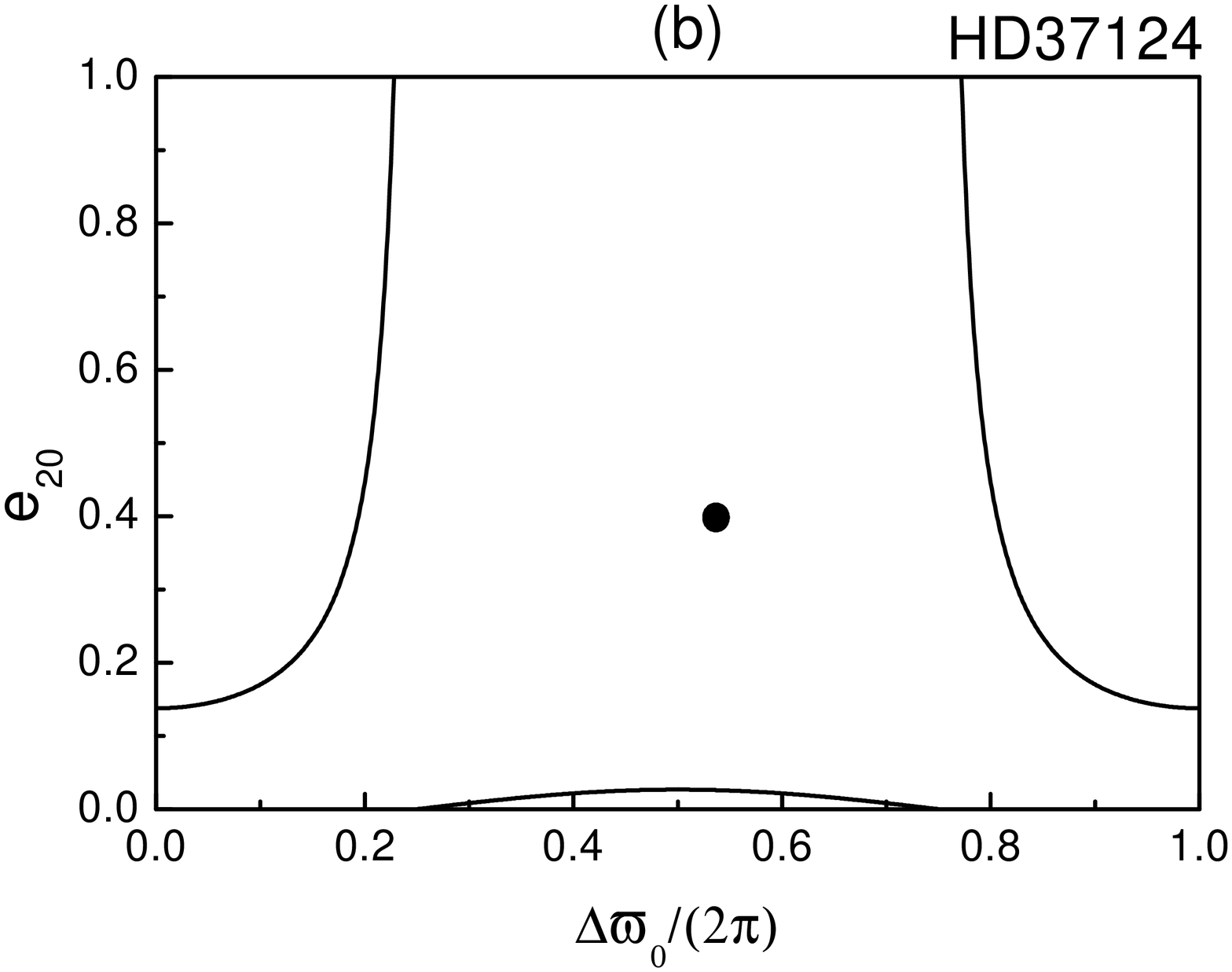}
\caption{
Libration region in the initial $e_{20}-\dw_{0}$ plane defined by (\ref{a17-1}) and (\ref{a17-2}) for
the planetary system (a) GJ876 ($\varpi_b=330^0$,$\varpi_c=333^0$),HD82943 ($\varpi_b=96^0$,$\varpi_c=138^0$);
(b) HD37124 ($\varpi_b=60^0$,$\varpi_c=259^0$).
The black dots show the present configuration of the two planets.
 Other orbital elements  are taken from Fischer et al. (2003).
 \label{fig13}}
\end{figure}

\clearpage

\begin{deluxetable}{lcc}
\tabletypesize{\scriptsize} \tablecaption{Orbital
Parameters\tablenotemark{a}~ for HD12661(1.07M$_{\bigodot}$) system \label{tb1}}
\tablewidth{0pt} \tablehead{ \colhead{Parameter} &
\colhead{HD12661 b} & \colhead{HD12661 c}} \startdata
planet mass M$\sin i$ (M$_{jup}$)  & 2.30 & 1.57 \\
period P(days) & 263.6(1.2) & 1444.5(12.5) \\
T$_p$ (JD) (days) & 2,449,941.9(6.2) & 2,449,733.6(49.0) \\
semi-major axis a (AU) & 0.82 & 2.56 \\
eccentricity e & 0.35(0.03) & 0.20(0.04)\\
argument of pericenter $\omega$ (deg)  & 293.1(5.0) & 162.4(18.5) \\
\enddata
\tablenotetext{a}{Data from  Fischer et al. (2003)}  
\end{deluxetable}

\begin{deluxetable}{lcc}
\tabletypesize{\scriptsize} \tablecaption{Orbital
Parameters\tablenotemark{b}~ for 47UMa(1.03M$_{\bigodot}$) system \label{tb2}}
\tablewidth{0pt} \tablehead{ \colhead{Parameter} & \colhead{47UMa
b}   & \colhead{47 UMa c}} \startdata
planet mass M$\sin i$ (M$_{jup}$)  & 2.54 & 0.76 \\
Period P(days) & 1089.0(2.9)&  2594(90)\\
T$_p$ (JD) (days) & 2,450,356.0(33.6) & 2,451,363.5(495.3) \\
semi-major axis a (AU) & 2.09 & 3.73 \\
eccentricity e & 0.061(0.014) & 0.005(0.115) \\
argument of pericenter $\omega$ (deg)  & 171.8(15.2) & 127.0(55.8)\\
\enddata
\tablenotetext{b}{Data from  Fischer et al. (2002)} 
\end{deluxetable}

\begin{deluxetable}{lccccc}
\tabletypesize{\scriptsize}
\tablecaption{Extensions of $\dw_0$ in apsidal resonance for the observed systems  \label{tb3}}
\tablewidth{0pt} \tablehead{ \colhead{Planet Pair} & \colhead{$q=\frac{m_1}{m_2}$}  & \colhead{$\alp=\frac{a_1}{a_2}$} &
\colhead{$\frac{e_{10}}{e_{20}}$} & \colhead{aligned $\dw_0$ } &
 \colhead{anti-aligned $\dw_0$} }
\startdata
Ups And b-c &  0.358& 0.0720 & 0.037  & (-79.3$^o$,79.3$^o$) & -\tablenotemark{c} \\
Ups And b-d &  0.181& 0.0228 & 0.040  & (-47.0$^o$,47.0$^o$) & -  \\
Ups And c-d &  0.507& 0.317 & 1.08  & (-9.0$^o$,9.0$^o$) & -  \\
55 Cnc b-c &  4.15& 0.477 & 0.073  & - & (96.8$^o$,263.2$^o$)  \\
55 Cnc b-d &  0.225 & 0.021 & 0.107  & - & -  \\
55 Cnc c-d &  0.054 & 0.044 & 1.46  & - & -  \\
GJ876 c-b &  0.296 & 0.628 & 2.70  & - & (144.0$^o$,216.0$^o$)  \\
47 UMa\tablenotemark{d} ~b-c &  3.34& 0.560 & 12.2  & (-87.9$^o$,87.9$^o$) & -  \\
HD37124 b-c & 0.860 & 0.184 & 0.250 & (-69.8$^o$,69.8$^o$) & - \\
HD12661 b-c & 1.46 & 0.320 & 1.75 & (-67.8$^o$,67.8$^o$) & (98.6$^o$,261.4$^o$) \\
HD82943 c-b & 0.540 & 0.628 & 1.32 & (-59.7$^o$,59.7$^o$) & (132.8$^o$,227.2$^o$) \\
HD168443 b-c &  0.450 & 0.103 & 2.65  & - & -  \\
HD38529 b-c & 0.061 & 0.035 & 0.806 & - & - \\
HD74156\tablenotemark{e} ~b-c & 0.208 & 0.080 & 1.625 & - & - \\
\enddata
\tablenotetext{c}{Here - means no possible libration $\dw_0$. }
\tablenotetext{d}{Data from  Fischer et al. (2002)} 
\tablenotetext{e}{Data from California and Carnegie Planet Search (2003)} 
\end{deluxetable}

\end{document}